\documentclass{article}

\usepackage{arxiv}
\usepackage[utf8]{inputenc}
\usepackage[T1]{fontenc}
\usepackage{microtype}
\usepackage{url}
\usepackage{graphicx}
\usepackage{float}
\usepackage{booktabs}
\usepackage{placeins}
\usepackage{wrapfig}
\usepackage{natbib}
\usepackage[hidelinks]{hyperref}

\usepackage{amsmath,amsfonts,bm}

\def\eqref#1{equation~\ref{#1}}

\def\1{\bm{1}}

\DeclareMathAlphabet{\mathsfit}{\encodingdefault}{\sfdefault}{m}{sl}
\SetMathAlphabet{\mathsfit}{bold}{\encodingdefault}{\sfdefault}{bx}{n}

\title{RePrior: Measurement-Reconciled Prior Correction\\for Sparse-View CBCT Reconstruction}
\date{}
\renewcommand{\headeright}{Preprint}
\renewcommand{\undertitle}{}
\renewcommand{\shorttitle}{RePrior}

\author{%
  \normalfont
  \begin{minipage}[t]{0.95\textwidth}
    \centering
    \normalsize
    Mengrui Yin\textsuperscript{1,2},
    Tianxingjian Ding\textsuperscript{1},
    Wufan Zhao\textsuperscript{2},\\[0.2em]
    Ziyang Yan\textsuperscript{3,4,*},
    Yu Tian\textsuperscript{1}\\[0.8em]
    {\small
      \textsuperscript{1} University of Central Florida\\[0.1em]
      \textsuperscript{2} The Hong Kong University of Science and Technology (Guangzhou)\\[0.1em]
      \textsuperscript{3} Tampere University\\[0.1em]
      \textsuperscript{4} Ellis Institute Finland\\[0.3em]
      \textsuperscript{*} Corresponding author: Ziyang Yan
    }
  \end{minipage}%
}

\hypersetup{
  pdftitle={RePrior: Measurement-Reconciled Prior Correction for Sparse-View CBCT Reconstruction},
  pdfauthor={Mengrui Yin, Tianxingjian Ding, Wufan Zhao, Ziyang Yan, Yu Tian},
  pdfsubject={Sparse-view cone-beam computed tomography reconstruction preprint},
  pdfkeywords={RePrior, CBCT, sparse-view reconstruction, Gaussian splatting, X-ray tomography}
}

\begin{document}
\maketitle

\begin{abstract}

Sparse-view cone-beam computed tomography (CBCT) is severely under-constrained by limited X-ray projections. Learned reconstruction can provide complementary anatomical information, but using it as a prior in patient-specific Gaussian reconstruction introduces two challenges. First, the discrepancy between the learned prior and current reconstruction is spatially non-uniform: some local differences contain useful missing 3D information, whereas others reflect prior bias. Second, even a volumetrically beneficial correction can conflict with the observed X-ray measurements. Since X-ray measurements are line integrals, spatially separated corrections may affect overlapping rays, with their signed projection responses reinforcing or partially canceling in the measurement residual, so their feasibility cannot always be assessed independently. In this paper, we introduce RePrior, a multi-stage progressive prior-correction framework that addresses correction actionability, measurement feasibility, and representation expressivity. Selective Correction Proposal (SCP) first identifies actionable signed local corrections from prior--reconstruction discrepancies. Measurement-Constrained Reconciliation (MCR) then jointly enforces their consistency with patient-specific X-ray measurements, while Hybrid Residual Enhancement (HRE) recovers additional correction capacity through complementary voxel- and Gaussian-domain degrees of freedom around the reconciled anchor. RePrior requires neither test-time ground truth nor case-specific network fine-tuning. RePrior gains approximately 2.6~dB in 3D PSNR over R$^2$-Gaussian across LUNA16, PANORAMA, and PENGWIN. Beyond prior-assisted reconstruction, it delivers additional gains in volumetric quality and observed-view projection fidelity, with consistent gains in held-out-view fidelity in the backbone-matched evaluation.
\end{abstract}
\section{Introduction}
\label{sec:introduction}

Cone-beam computed tomography (CBCT) reconstructs a three-dimensional attenuation field from multi-view X-ray projections.
Reducing the number of acquired views is a direct way to lower radiation exposure and acquisition burden~\citep{sidky2006fewview,bian2010sparseview}. This is particularly attractive in settings
where CBCT is acquired repeatedly or under tight acquisition constraints. However, reduced angular sampling can produce prominent streak artifacts and degrade the recovery of structural details in reconstructed CT images
\citep{bian2010sparseview}. Preserving these local anatomical details is important for
anatomically faithful 3D reconstruction and interpretation~\citep{lin2025pixel}.
Early learning-based approaches mitigated these sparse-view artifacts through image-domain or dual-domain reconstruction networks
\citep{jin2017fbpconvnet,chen2018learn,zhang2018sparse,wu2021drone}.
More recently, neural fields and explicit Gaussian representations have enabled
patient-specific 3D reconstruction directly from measured X-rays through
differentiable projection
\citep{zha2022naf,cai2024sax-nerf,zha2024r2-gs,
yang2026exact-gs}.
Complementary directions include reconstruction-friendly projection interpolation~\citep{wang2022rfinet}, meta-learned neural attenuation fields~\citep{shin2025fact}, learned 3D Gaussian feature representations~\citep{lin2024difgaussian}, and diffusion priors coupled with explicit measurement models~\citep{liu2023dolce}.
Despite improved representations and optimization strategies, highly sparse measurements remain challenging for 3D reconstruction.

\begin{figure}[htbp]
    \centering
    \includegraphics[width=0.98\linewidth]
    {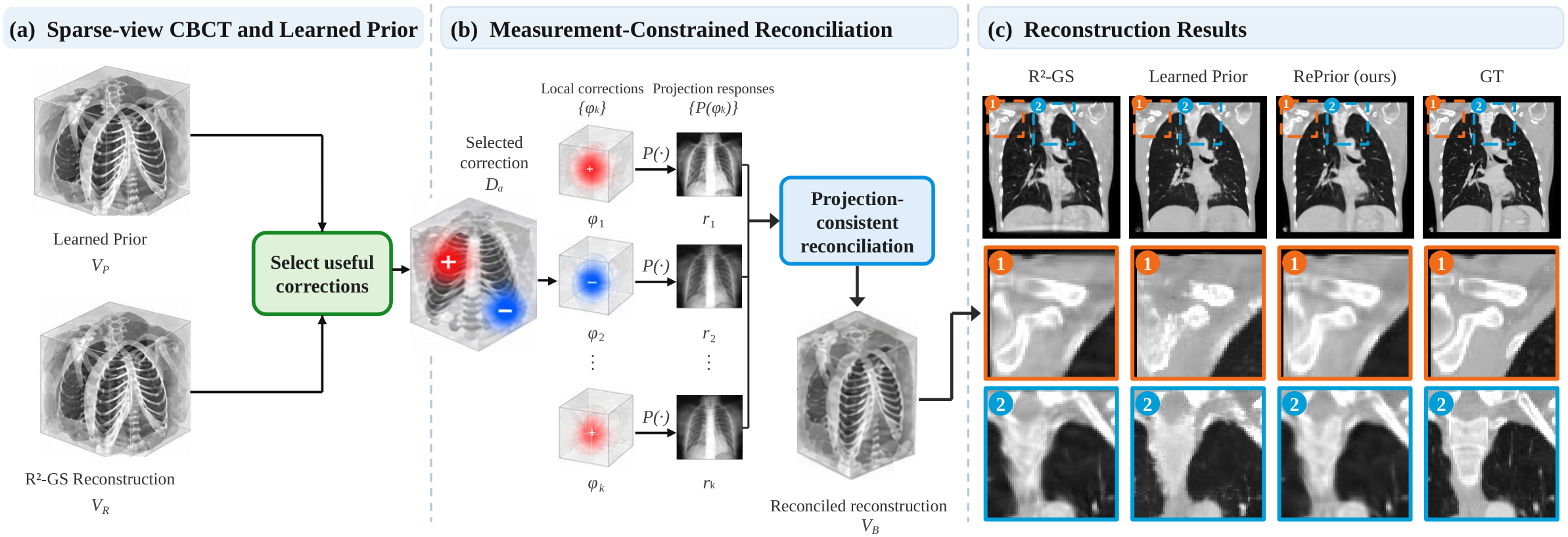}
    \caption{
    We introduce \textbf{RePrior}, a  multi-stage progressive prior-correction framework which transfers learned priors into patient-specific sparse-view CBCT
    reconstruction through selective correction proposal, measurement-constrained
    reconciliation, and hybrid residual enhancement.
    }
    \label{fig:teaser}
\end{figure}

For sparse-view CBCT, the central challenge is not merely to recover
plausible anatomy, but to recover anatomy that is supported by the measurements
of the current patient. To complement the weakly constrained 3D information,
recent feed-forward approaches such as X-GRM~\citep{liu2025X-GRM},
X-LRM~\citep{zhang2026X-lrm}, and DeepSparse~\citep{lin2026deepsparse}
learn mappings from sparse X-rays to volumetric or explicit 3D
representations, providing strong anatomical information learned from training
data. Here, we focus on a complementary post-reconstruction question: whether individual local prior-induced changes are supported by the current patient's measured X-ray projections. Complementarily, patient-specific neural-field and Gaussian reconstruction methods directly optimize against the
current measurements and therefore provide strong data fidelity, yet sparse
acquisition still leaves substantial 3D degrees of freedom weakly constrained.
This raises a key question: how can complementary 3D information from a learned reconstruction be converted into patient-specific corrections that are supported by the measured X-rays, rather than simply choosing between a learned prior and projection fitting?

\begin{wrapfigure}{r}{0.65\textwidth}
    \vspace{-1em}
    \centering
    \includegraphics[width=\linewidth]
    {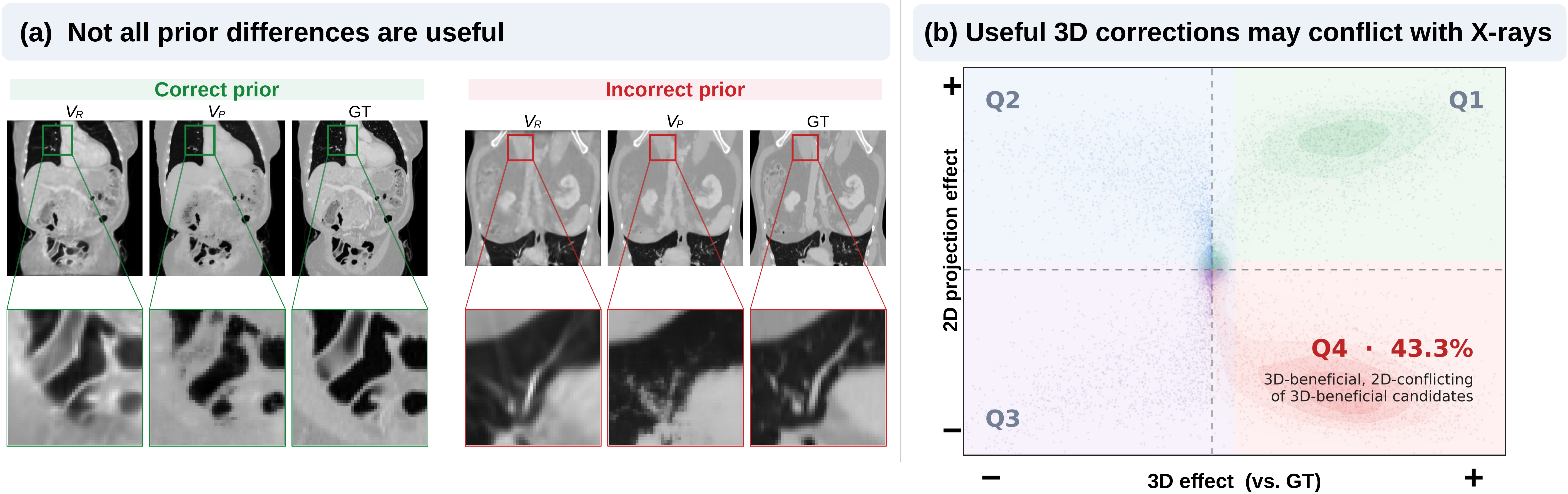}
    \vspace{-2em}
    \caption{
    Challenges in prior-guided correction. Left: local prior differences contain both useful 3D corrections
    and prior bias. Right: some volumetrically beneficial corrections still conflict with X-ray measurements.
    }
    \label{fig:motivation}
    \vspace{-1em}
\end{wrapfigure}

Figure~\ref{fig:motivation} further reveals two recurring difficulties across
the three anatomies studied here. First, local prior reliability is spatially
non-uniform: every analyzed case contains regions where $V_P$ is closer to the
reference volume than $V_R$, as well as regions where it is farther away,
indicating that the remaining discrepancy mixes useful 3D information with
local prior bias. Second, volumetric benefit does not guarantee measurement
compatibility: our post-hoc diagnostic shows that 43.3\% of the 3D-beneficial
recovery candidates increase the projection error over the observed X-ray
views, with such conflicts occurring in all 230 analyzed cases. These
observations lead to two sequential questions: which local prior differences
should be adopted, and how should the selected corrections be jointly
reconciled under patient-specific X-ray measurements?


Based on these observations, we propose RePrior, a multi-stage
progressive prior-correction framework for sparse-view CBCT.
Fig.~\ref{fig:teaser} summarizes the overall process of prior-to-reconstruction
correction-transfer.
Rather than relying on direct global prior supervision alone, RePrior further
decomposes prior correction into three progressive stages:
correction actionability, measurement feasibility, and representation expressivity.
First, Selective Correction Proposal (SCP) learns which local prior differences are actionable and how strongly their signed discrepancy directions should be retained.
Next, Measurement-Constrained Reconciliation (MCR) coordinates multiple candidate corrections according to their patient-specific X-ray responses, thereby establishing a measurement-feasible reconstruction.
Finally, Hybrid Residual Enhancement (HRE) uses this feasible solution as an anchor and jointly exploits voxel- and Gaussian-domain residual freedom to recover useful 3D corrections that remain unexpressed after reconciliation.

In this paper, our contributions are summarized as follows:
\begin{itemize}
    \item
    We propose RePrior, a progressive prior-correction framework
    for sparse-view CBCT that transfers learned volumetric priors into
    patient-specific Gaussian reconstruction through a three-stage progression
    from correction actionability to measurement feasibility and representation
    expressivity, rather than direct global prior fusion.
    \item
We conduct a systematic analysis of prior correction in sparse-view CBCT and identify two key failure modes: local prior discrepancies have spatially non-uniform utility, and volumetrically beneficial corrections do not necessarily improve patient-specific X-ray fitting. These findings motivate separating selective correction proposal from measurement-constrained reconciliation.   
    \item
    Across three anatomies and 230 evaluation cases, RePrior improves 3D PSNR by approximately 2.6 dB over R$^2$-Gaussian. Beyond the matched prior-assisted reconstruction, its correction stages further improve volumetric quality and observed- and held-out-view projection fidelity.
\end{itemize}
\section{Related Work}
\label{sec:related_work}

\subsection{Sparse-View Tomographic Reconstruction}

Sparse-view reconstruction has attracted considerable attention in medical CT
imaging.
Early deep-learning methods~\citep{zhang2018sparse,wu2021drone}
refined analytical reconstructions to suppress undersampling artifacts, while
neural fields enabled direct optimization against measured projections.
NAF~\citep{zha2022naf} models attenuation with a coordinate-based network
through differentiable X-ray rendering, and
X-Gaussian~\citep{cai2024radiative} adapts 3D Gaussian
Splatting~\citep{kerbl20233d} for efficient X-ray optimization and rendering.

More recently, feed-forward models such as
X-GRM~\citep{liu2025X-GRM},
X-LRM~\citep{zhang2026X-lrm}, and
DeepSparse~\citep{lin2026deepsparse}
learn transferable 3D reconstruction priors from large-scale data.
Generative approaches such as CvG-Diff~\citep{chen2025cvgdiff} and
CT-SDM~\citep{yang2025ctsdm} further exploit diffusion-based priors to recover
anatomical structures under severe angular undersampling.
These learned reconstruction models provide strong structural information.
RePrior focuses on a complementary question: which remaining local
prior-induced changes should be retained under the measured projections of
the current patient.

\subsection{Tomographic Reconstruction with 3DGS Representation}

Although early NeRF-based approaches~\citep{zha2022naf,cai2024sax-nerf}
enable high-fidelity tomographic reconstruction, their implicit representations
require per-scene optimization.
R$^2$-Gaussian~\citep{zha2024r2-gs} introduced 3DGS-based sparse-view CT
reconstruction with X-ray-specific rasterization for efficient reconstruction.

Subsequent works advance Gaussian tomography from complementary perspectives.
3DGR-CT~\citep{li2025_3dgrct} combines FBP-guided Gaussian initialization
with a differentiable CT projector, while
Exact-GS~\citep{yang2026exact-gs} derives an exact splatting formulation for
X-ray reconstruction.
PGE~\citep{wu2026pge} further explores uncertainty-guided active view
selection.

These advances establish 3D Gaussians as an efficient explicit representation
for projection-driven tomographic reconstruction. RePrior studies a
complementary problem: selectively transferring learned volumetric prior
corrections into a patient-specific Gaussian reconstruction.

\subsection{Learned Priors with Measurement Constraints}

Another line of work combines learned generative priors with explicit physical
measurements during reconstruction~\citep{chung2023solving}.
DOLCE~\citep{liu2023dolce} couples a conditional diffusion prior with an
explicit CT measurement model for limited-angle reconstruction.
Recent work has also studied discrepancies introduced by data-consistency
operations in diffusion-based sparse-view CT reconstruction~\citep{chen2025cddm},
while diffusion generative priors have been incorporated into iterative
reconstruction through physics-informed initialization and latent
optimization~\citep{evangelista2026diffusionprior}.
DiffNR~\citep{su2026diffnr} instead uses diffusion-generated pseudo-reference
volumes to provide auxiliary 3D supervision during neural representation
optimization for sparse-view tomography.

These studies demonstrate the value of combining learned structural priors
with physical measurements for tomographic reconstruction.
RePrior builds on this general direction and focuses on selectively utilizing
the residual prior discrepancies that remain after prior-assisted Gaussian
reconstruction, with subsequent correction constrained by patient-specific
X-ray measurements.
\section{Problem Analysis: Prior Utility and Measurement Compatibility}
\label{sec:problem_analysis}

Learned-prior utilization in sparse-view CBCT faces two related but distinct
challenges. Let $V_P$ denote the learned prior and $V_R$ the corresponding
prior-assisted Gaussian reconstruction, as illustrated in
Fig.~\ref{fig:motivation}; their formal construction is introduced in
Sec.~\ref{sec:method_overview}.

First, the remaining prior discrepancy does not have uniformly reliable local
utility. Sparse-view measurements may admit multiple attenuation volumes
consistent with the same data~\citep{sidky2006fewview,bian2010sparseview}.
This motivates our use of a learned prior as a source of complementary
structural information. However, this does not imply that the remaining
discrepancy $V_P-V_R$ should be transferred unconditionally. As
shown in Fig.~\ref{fig:motivation}(a), within the same patient, some regions of
the learned prior are closer to the reference volume, whereas others exhibit
larger prior bias. The remaining discrepancy should therefore be regarded as
candidate correction information with unknown local reliability. Reliable
prior correction must first determine which local differences correspond to
useful structures still missing from the current reconstruction and how
strongly they should be retained.

Second, volumetric utility does not imply measurement compatibility. As shown
in Fig.~\ref{fig:motivation}(b), even a local change that improves the
volumetric reconstruction can worsen the fit to the patient's observed X-ray
measurements. Our post-hoc diagnostic shows that this conflict is systematic:
43.3\% of the 3D-beneficial recovery candidates increase the projection error
over the observed X-ray views, and such conflicting corrections occur in all
230 analyzed cases. Thus, potential volumetric benefit alone is insufficient
for deciding whether a correction should be retained.

Moreover, under the calibrated log-domain line-integral model, the responses
of multiple local corrections add in projection space. Let
$r=\mathcal{P}(V_R)-Y$ denote the current measurement residual and
$h_i=\mathcal{P}(\delta V_i)$ the projection response of correction
$\delta V_i$. For two corrections, the change in squared projection error is
\begin{equation}
\begin{aligned}
&\|r+h_i+h_j\|_2^2-\|r\|_2^2 \\
&=2\langle r,h_i+h_j\rangle
+\|h_i\|_2^2+\|h_j\|_2^2
+2\langle h_i,h_j\rangle .
\end{aligned}
\label{eq:joint_projection_error}
\end{equation}
The cross term $2\langle h_i,h_j\rangle$ shows that the measurement effect of
one correction depends on the other corrections applied simultaneously:
their signed ray responses may reinforce or partially cancel one another.
Thus, measurement compatibility cannot in general be determined by accepting
or rejecting local corrections independently.

Reliable prior correction therefore requires answering two distinct questions
in sequence:
which local prior differences should be adopted, and how should the resulting
candidate changes remain compatible with the current patient's X-ray
measurements when acting jointly?
The former concerns the local reliability and utility of prior information,
whereas the latter requires reasoning about the joint ray effects of multiple
corrections in patient-specific measurement space.
\section{Method}
\label{sec:method}

\subsection{Overview}
\label{sec:method_overview}

RePrior addresses the requirements identified in Sec.~\ref{sec:problem_analysis}
by selecting useful prior differences and reconciling their joint X-ray effects. Given sparse-view X-ray measurements $\mathcal{Y}$, a pretrained
reconstructor $F_P$ first produces a learned volumetric prior $V_P$.
We additionally construct a patient-specific prior-assisted Gaussian
reconstruction by optimizing R$^2$-GS against the measured projections and
subsequently introducing fixed spatial supervision from $V_P$.
This yields the Gaussian representation $\mathcal{G}_R$ and its voxelized
attenuation volume $V_R$:
\begin{equation}
    V_P = F_P(\mathcal{Y}), \qquad
    \mathcal{G}_R =
    \mathcal{R}_{\mathrm{R^2GS}}(\mathcal{Y};V_P),
    \qquad
    V_R = \mathcal{V}(\mathcal{G}_R),
    \label{eq:initial_recon}
\end{equation}
where $\mathcal{V}(\cdot)$ denotes Gaussian voxelization.
The detailed optimization schedule used to construct $V_R$ is provided in
the Appendix.

\begin{figure}[t]
    \centering
    \includegraphics[width=0.92\linewidth]
    {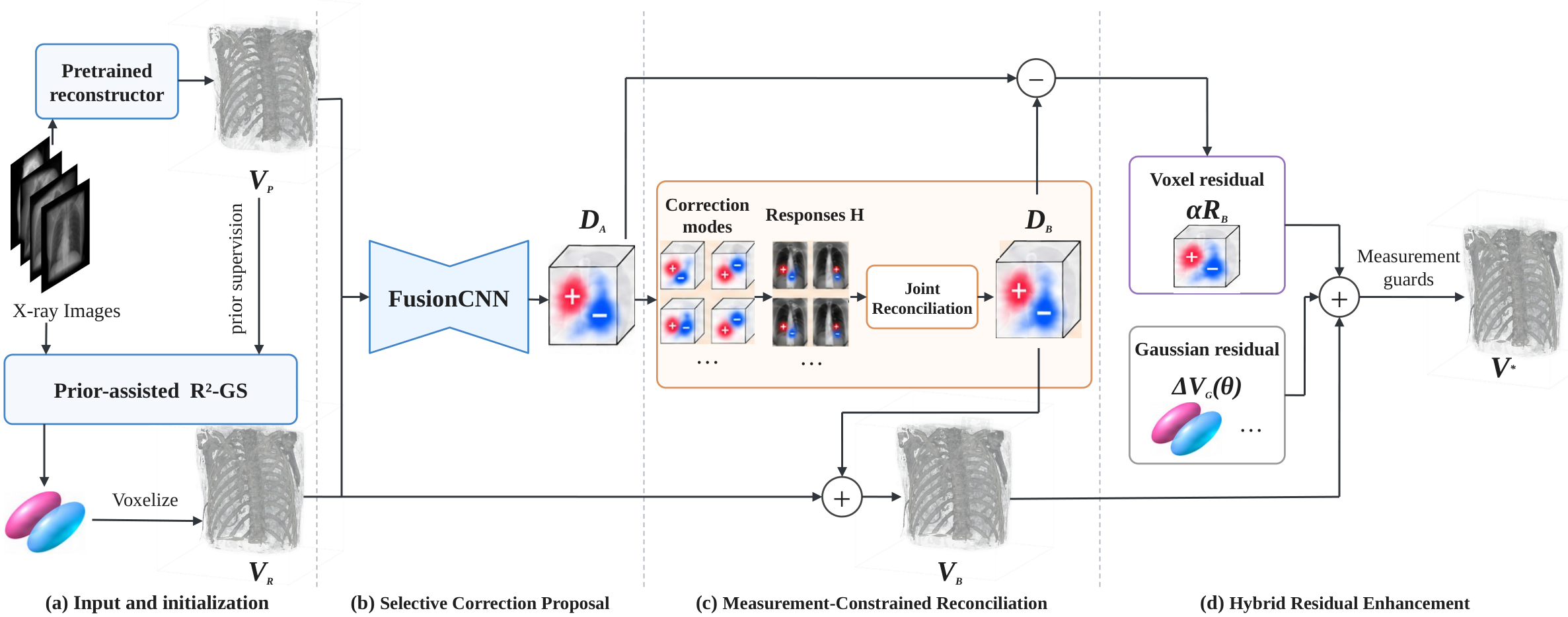}
    \caption{
    Overview of RePrior.
    (a) Learned prior $V_P$ and prior-assisted reconstruction $V_R$.
    (b) SCP predicts a signed correction $D_A$.
    (c) MCR jointly reconciles structured correction modes using their
    X-ray responses $H$, producing $D_B$ and the feasible anchor $V_B$.
    (d) HRE combines voxel- and Gaussian-domain residual freedom under
    measurement guards to obtain $V^*$.
    }
    \label{fig:reprior_overview}
\end{figure}

As illustrated in Fig.~\ref{fig:reprior_overview}, SCP converts the remaining
prior discrepancy into an actionable signed correction $D_A$.
MCR jointly reconciles its patient-specific X-ray responses, yielding
$D_B$ and the measurement-feasible anchor $V_B$.
Because MCR uses a restricted structured correction space, HRE introduces
complementary voxel- and Gaussian-domain residual freedom around $V_B$
to recover unexpressed SCP corrections while preserving measurement consistency.

\subsection{Selective Correction Proposal}
\label{sec:scp}

The discrepancy $V_P-V_R$ should be regarded as candidate prior information rather than a direct correction target. As analyzed in Sec.~\ref{sec:problem_analysis}, such local disagreement may correspond either to anatomical structure genuinely missed under sparse acquisition or to prediction bias introduced by the learned reconstructor. Its magnitude alone does not indicate local prior reliability.

To distinguish these cases, SCP uses training-time ground-truth supervision
to learn which observable prior--reconstruction discrepancies are actionable
and how strongly their signed directions should be retained.
We collectively denote local frequency characteristics, boundary and
differential morphology, and structural disagreement cues by
$\Phi(V_P,V_R)$.
A FusionCNN estimates the local actionability and retained magnitude of the
prior discrepancy, yielding
\begin{equation}
D_A =
F_{\mathrm{SCP}}
\big(
V_P,
V_R,
\Phi(V_P,V_R)
\big),
\label{eq:scp}
\end{equation}
where the spatial support and magnitude determine where and how strongly the
prior discrepancy is retained, while the attenuation-change direction follows
the sign of $V_P-V_R$.
SCP thus converts the spatially non-uniform prior discrepancy into an actionable
local correction proposal for subsequent measurement-constrained reconciliation.
\subsection{Measurement-Constrained Reconciliation}
\label{sec:mcr}

\begin{wrapfigure}{R}{0.52\textwidth}
    \centering
    \includegraphics[width=\linewidth]
    {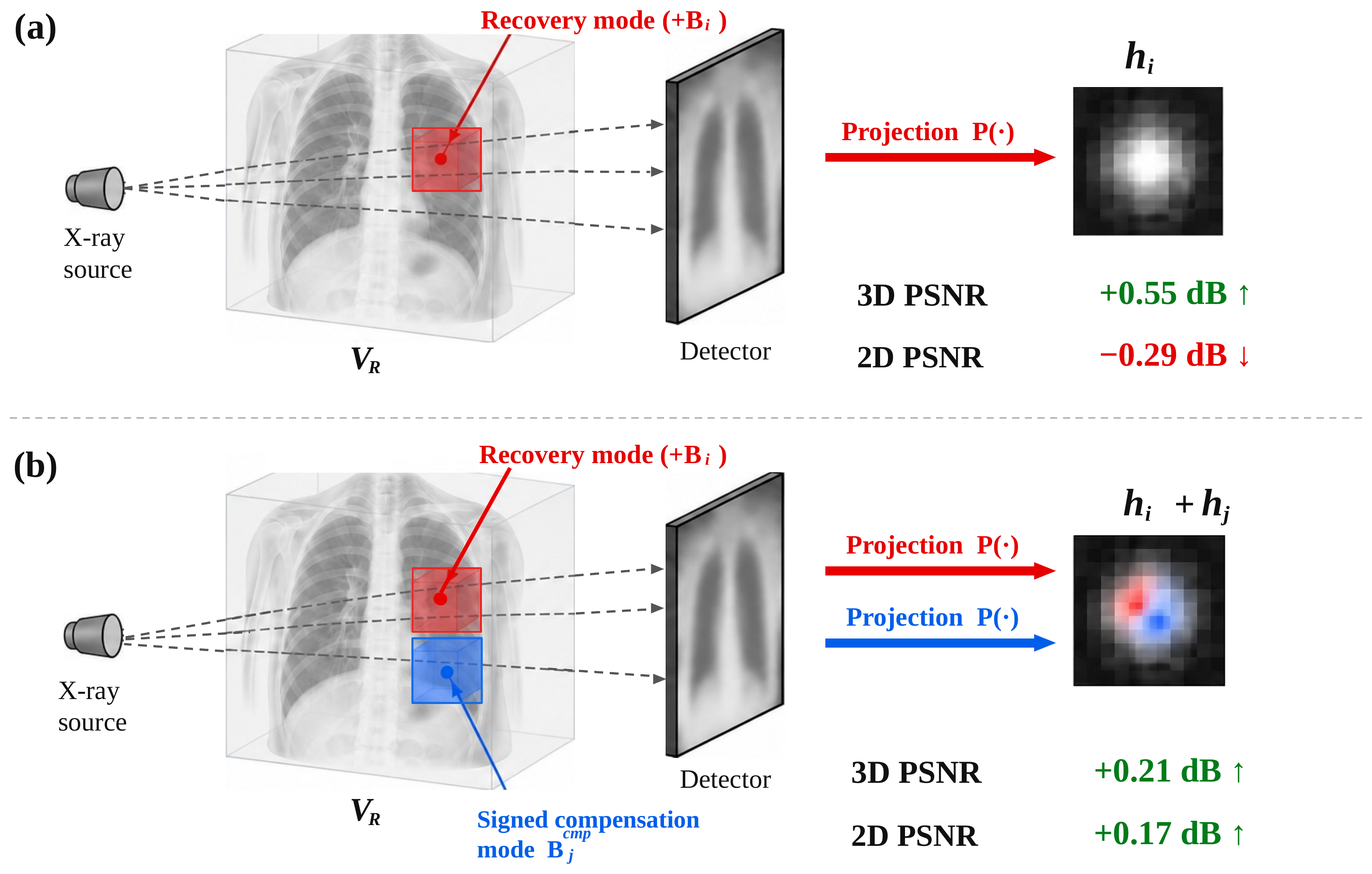}
    \vspace{-1em}
    \caption{Ray-coupled correction responses on a representative case, illustrating measurement conflict and signed compensation.}
    \label{fig:mcr_joint_feasibility}
\end{wrapfigure}

SCP identifies locally beneficial corrections in the volumetric domain, but
volumetric utility does not guarantee consistency with the measured X-rays.
As shown in Fig.~\ref{fig:motivation}(b), some corrections that improve the
volume can still degrade the X-ray fitting of the current patient.
Since each X-ray measurement integrates attenuation along a source--detector
path, spatially separated corrections may affect the same measurements.
As illustrated in Fig.~\ref{fig:mcr_joint_feasibility}, a correction that is
beneficial in 3D can worsen projection fitting when considered alone, while
another signed mode can partially compensate for its ray response.
This motivates reconciling multiple corrections jointly in measurement space.

MCR organizes $D_A$ into a low-dimensional structured correction space.
We first establish a coarse measurement-aware correction and then parameterize
the remaining correction using regional recovery modes, which control how much
of the desirable correction is retained across spatial regions.
However, resolving measurement conflicts through suppression alone may discard
useful 3D changes.
We therefore introduce signed compensation modes that provide additional
attenuation-redistribution freedom across regions, allowing their measurement
responses to reinforce or partially cancel rather than simply suppressing the
original correction.\footnote{Detailed construction of the coarse correction,
regional recovery modes, and signed compensation modes is provided in the
Appendix.}

We then map the variable correction modes into the X-ray measurement space of
the current patient.
Rather than reasoning about candidate corrections independently in voxel space,
their patient-specific ray responses directly characterize how each 3D change
perturbs the measured projections.
The corresponding coefficients are jointly optimized to preserve as much of the
SCP proposal as possible while satisfying patient-specific measurement and
physical constraints:

\begin{equation}
\mathbf{c}^{*}
=
\arg\min_{\mathbf{c}\in\mathcal{C}_{\mathrm{phys}}}
\left\|
D_B(\mathbf{c})-D_A
\right\|_2^2
\quad
\mathrm{s.t.}\quad
\mathcal{C}_{\mathrm{meas}}
\big(
r_0-H\mathbf{c}
\big)
\leq 0 .
\label{eq:mcr_objective}
\end{equation}
Here, $D_B(\mathbf{c})$ denotes the correction synthesized from the coarse
measurement-aware correction and the variable recovery and compensation modes,
$H$ collects the patient-specific X-ray responses of the variable modes, and
$r_0$ denotes the measurement residual after the coarse correction.
The optimized correction yields the reconciled reconstruction
$V_B=V_R+D_B(\mathbf{c}^{*})$, which serves as the measurement-feasible anchor
for HRE.

\subsection{Hybrid Residual Enhancement}
\label{sec:hre}


Since MCR operates in a restricted structured correction space
to obtain a controllable measurement-feasible solution, it may not realize
the entire SCP proposal.
The unrealized component does not necessarily correspond to incorrect prior
information, but may instead arise from the limited expressivity of the MCR
correction modes.
We denote this residual correction by
$R_B=D_A-D_B$ and make no assumption that $R_B$ is measurement-feasible on
its own.

We then introduce HRE. After MCR establishes a measurement-feasible anchor $V_B$, HRE recovers additional representation freedom around this anchor to capture useful corrections that cannot be represented within the restricted MCR space, while preserving the measurement feasibility established by MCR.

Specifically, HRE parameterizes this residual search using two complementary
degrees of freedom.
The voxel branch moves along the unresolved SCP direction $R_B$ through a
scalar coefficient $\alpha$, preserving its signed attenuation structure.
In parallel, the Gaussian branch starts from the fixed-topology
prior-assisted Gaussian state
$\mathcal{G}_{\mathrm{ref}}=\mathcal{G}_R$
and performs bounded residual updates to attenuation, position, scale, and
orientation, enabling geometric and spatial-support redistribution that
cannot be expressed by voxel-domain backfilling alone.
Together, the two branches define the candidate reconstruction
\begin{equation}
    V_C(\alpha,\theta)
    =
    V_B
    +
    \alpha R_B
    +
    \left[
        \mathcal{V}(\mathcal{G}_{\theta})
        -
        \mathcal{V}(\mathcal{G}_{\mathrm{ref}})
    \right].
\label{eq:joint_hre}
\end{equation}


Here, $\alpha R_B$ represents voxel-domain residual backfilling, while the Gaussian term captures the volumetric correction induced by the bounded Gaussian updates. HRE jointly optimizes these two residual components and retains only candidates that satisfy the measurement guards\footnote{More details are provided in the Appendix.}. The selected candidate is denoted as the final reconstruction $V^{*}$.

\FloatBarrier
\section{Experiments}
\label{sec:experiments}

\subsection{Experimental Setting}
\label{sec:exp_setup}

\noindent\emph{Datasets and evaluation.}
We evaluate on LUNA16~\citep{setio2017luna16},
PANORAMA~\citep{alves2026panorama}, and PENGWIN~\citep{sang2026pengwin}.
Unless otherwise specified, matched experiments use 25 observed projections
and 40 held-out views reserved for evaluation. 
We report volumetric PSNR/SSIM and reproject each reconstruction with the
same physics-based projector to compute observed- and held-out-view
PSNR/SSIM under a common per-view normalization protocol.
Reconstruction outputs are fixed before ground-truth and held-out evaluation.

\noindent\emph{Baselines and implementation.}
We compare with SAX-NeRF, X-Gaussian, R$^2$-GS, Exact-GS, and DeepSparse.
DeepSparse-10 is the official 10-view checkpoint, reported as an additional
reference; DeepSparse-25 is our continuation trained under the matched 25-view
setting. RePrior uses R$^2$-GS as its primary backbone and a frozen SCP
predictor at test time. Further training and implementation details are
provided in the Appendix.

\subsection{Main Comparison}

\label{sec:main_results}

\noindent\emph{Quantitative evaluation.}
As shown in Table~\ref{tab:main_results}, RePrior consistently achieves the
highest 3D PSNR and SSIM across all three evaluated anatomies.
Compared with R$^2$-GS, RePrior improves 3D PSNR by $+3.46$~dB on LUNA16,
$+2.38$~dB on PANORAMA, and $+2.04$~dB on PENGWIN, corresponding to a
dataset-level macro-average gain of approximately $2.63$~dB. Relative to the strongest competing method in each dataset, RePrior further
improves 3D PSNR by $+1.73$~dB on LUNA16, $+1.93$~dB on PANORAMA, and
$+1.47$~dB on PENGWIN.
The corresponding SSIM results show the same consistent trend.

These results demonstrate the complementary benefit of learned structural
information and patient-specific measurement constraints.
Learned reconstruction provides anatomical information that is difficult to
recover from sparse projections alone, while RePrior selectively transfers and
reconciles useful local corrections within a patient-specific reconstruction
rather than directly imposing the learned volume.


\begin{table*}[htbp]
\centering
\small
\setlength{\tabcolsep}{4.2pt}
\caption{
Main comparison on sparse-view CBCT reconstruction.
Best and second-best results are shown in bold and underlined, respectively.
}
\label{tab:main_results}

\begin{tabular}{@{}lcccccc@{}}
\toprule
&
\multicolumn{2}{c}{LUNA16 (90)}
&
\multicolumn{2}{c}{PANORAMA (90)}
&
\multicolumn{2}{c}{PENGWIN (50)}
\\
\cmidrule(lr){2-3}
\cmidrule(lr){4-5}
\cmidrule(lr){6-7}

Method
& PSNR$\uparrow$ & SSIM$\uparrow$
& PSNR$\uparrow$ & SSIM$\uparrow$
& PSNR$\uparrow$ & SSIM$\uparrow$
\\
\midrule

SAX-NeRF (CVPR'24)
& 29.41 & 0.8311
& 27.57 & 0.8358
& \underline{30.86} & 0.8726
\\

X-Gaussian (ECCV'24)
& 19.55 & 0.3030
& 19.24 & 0.3546
& 20.41 & 0.3468
\\

R$^2$-GS (NeurIPS'24)
& 29.12 & 0.8357
& 28.42 & 0.8534
& 30.29 & \underline{0.8817}
\\

Exact-GS (CVPR'26)
& 26.71 & 0.7598
& 26.51 & 0.7711
& 28.46 & 0.8134
\\

DeepSparse-10 (TMI'26)
& 30.74 & 0.8660
& 26.35 & 0.8386
& 27.12 & 0.8461
\\

DeepSparse-25
& \underline{30.85} & \underline{0.8750}
& \underline{28.87} & \underline{0.8780}
& 29.33 & 0.8740
\\

\midrule

\textbf{RePrior (Ours)}
& \textbf{32.58} & \textbf{0.8881}
& \textbf{30.80} & \textbf{0.8990}
& \textbf{32.33} & \textbf{0.9146}
\\

\bottomrule
\end{tabular}
\vspace{-1em}
\end{table*}

\begin{figure*}[t]
    \centering
    \includegraphics[width=\textwidth]
    {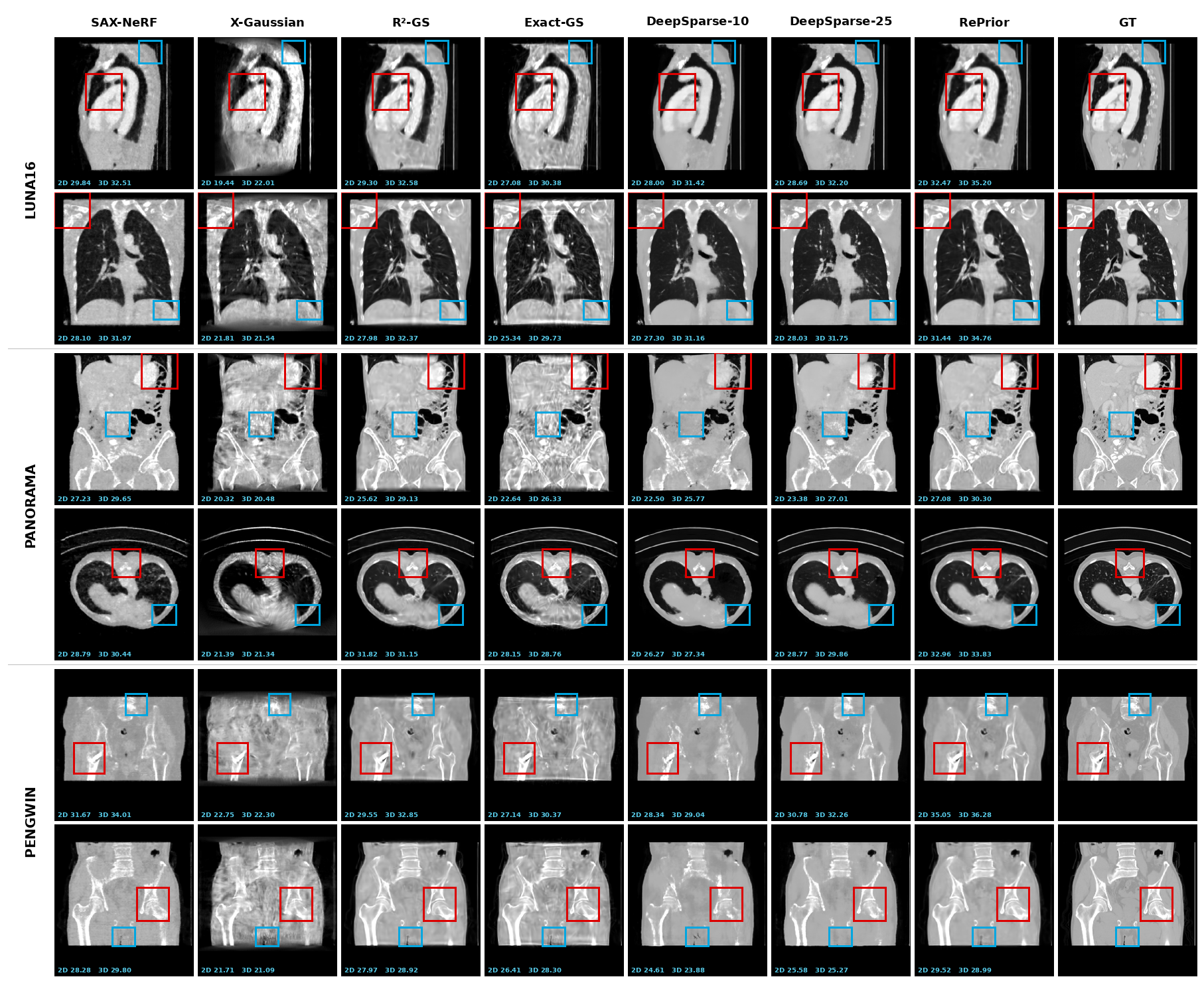}

    \vspace{-1em}
    \caption{
    Qualitative comparison on representative LUNA16, PANORAMA, and PENGWIN cases.
    Red and cyan boxes mark the highlighted local regions.
    }
    \vspace{-1.5em}\label{fig:qualitative_results}
\end{figure*}

\noindent\emph{Qualitative evaluation.}
Figure~\ref{fig:qualitative_results} shows representative reconstructions
across the three anatomies.
Compared with the projection-driven Gaussian baselines, RePrior better
preserves the highlighted local boundaries and fine structures while reducing
local blur and structural distortion.
The DeepSparse variants recover plausible global anatomy but show local
smoothing or attenuation mismatch in several highlighted regions; RePrior
more closely follows the ground-truth local shape and attenuation pattern.
These patterns are visible in the representative thoracic structures on LUNA16,
abdominal and pelvic interfaces on PANORAMA, and fine osseous structures on
PENGWIN, complementing the quantitative improvements in
Table~\ref{tab:main_results}.

\subsection{Generalization Across Gaussian Backbones}
\label{sec:backbone_generalization}

To examine whether RePrior generalizes across Gaussian reconstruction
backbones, we instantiate the same correction framework with both R$^2$-GS
and Exact-GS on LUNA16.
For each backbone, we report the vanilla reconstruction, its matched
prior-assisted reconstruction, and the final RePrior result.
The 3D metrics are volumetric PSNR/SSIM, while the 2D metrics are evaluated
on the 40 held-out projections.
The R$^2$-GS comparison uses 90 cases, whereas Exact-GS uses the available
30-case matched cohort; we therefore interpret the progression within each
backbone rather than comparing absolute values across backbones.

\begin{table}[!htbp]
\centering
\small
\setlength{\tabcolsep}{4.2pt}
\caption{
Generalization across Gaussian reconstruction backbones on LUNA16.
}
\label{tab:backbone_generalization}

\begin{tabular}{@{}lclcccc@{}}
\toprule
Backbone
& Cases
& Method
& 3D PSNR$\uparrow$
& 3D SSIM$\uparrow$
& 2D PSNR$\uparrow$
& 2D SSIM$\uparrow$
\\
\midrule

R$^2$-GS
& 90
& R$^2$-GS
& 29.12
& 0.8357
& 43.17
& 0.9864
\\

&
90
& Prior-R$^2$-GS
& 32.09
& 0.8826
& 44.09
& 0.9906
\\

&
90
& \textbf{RePrior}
& \textbf{32.58}
& \textbf{0.8881}
& \textbf{44.24}
& \textbf{0.9910}
\\

\midrule

Exact-GS
& 30
& Exact-GS
& 27.25
& 0.7749
& 40.60
& 0.9754
\\

&
30
& Prior-Exact-GS
& 31.27
& 0.8688
& 44.22
& 0.9893
\\

&
30
& \textbf{RePrior}
& \textbf{31.87}
& \textbf{0.8754}
& \textbf{44.35}
& \textbf{0.9898}
\\

\bottomrule
\end{tabular}
\end{table}

As shown in Table~\ref{tab:backbone_generalization}, RePrior provides
additional gains beyond the matched prior-assisted reconstruction for both
Gaussian backbones. With R$^2$-GS, RePrior improves 3D PSNR by
$+0.49$~dB over Prior-R$^2$-GS (32.09$\rightarrow$32.58~dB), while
held-out projection PSNR also increases from 44.09 to 44.24~dB.
With Exact-GS, the corresponding additional 3D gain is $+0.60$~dB
(31.27$\rightarrow$31.87~dB), with held-out projection PSNR increasing
from 44.22 to 44.35~dB. This consistent
Vanilla$\rightarrow$Prior-assisted$\rightarrow$RePrior progression indicates
that the proposed correction framework is not tied to a single Gaussian
backbone.
\subsection{Stage-wise Ablation and Analysis}
\label{sec:ablation}








The three stages of RePrior serve complementary rather than redundant roles.
SCP first proposes local corrections with potential volumetric utility, MCR
reconciles these corrections under patient-specific X-ray measurements, and
HRE restores residual expressivity around the resulting measurement-feasible anchor. Accordingly, these stages should not be interpreted as repeatedly optimizing the
same 3D objective, and their addition need not produce a monotonic increase in
3D PSNR.

Table~\ref{tab:stage_ablation} summarizes the stage-wise contribution of each
component in RePrior. For this ablation, we use 20 randomly selected cases from each dataset and
evaluate all configurations on the same cases. Results are reported
as dataset-balanced averages across the three datasets, and the 2D metrics
are computed on the 25 observed projections.

SCP recovers useful volumetric information from the prior--reconstruction
discrepancy, without explicitly enforcing measurement compatibility. Compared
with Prior-R$^2$-GS, adding SCP increases 3D PSNR from 32.03 to 32.58~dB
($+0.55$~dB) and 3D SSIM from 0.8966 to 0.9033. This gain comes with a slight
decrease in observed-view projection PSNR from 46.46 to 46.30~dB and SSIM from
0.9956 to 0.9952, confirming that volumetric improvement does not necessarily
imply measurement compatibility.

MCR then reconciles the candidate corrections under the patient-specific
measurements. From SCP to MCR, 3D PSNR decreases from 32.58 to 32.24~dB and
3D SSIM from 0.9033 to 0.8989, while observed-view projection PSNR increases
from 46.30 to 46.63~dB ($+0.33$~dB) and SSIM from 0.9952 to 0.9957.
Importantly, MCR still improves over Prior-R$^2$-GS in both 3D PSNR
(32.24 vs.\ 32.03~dB) and projection PSNR (46.63 vs.\ 46.46~dB), indicating
that measurement reconciliation preserves useful prior corrections rather
than simply removing them.

\begin{wraptable}{r}{0.52\textwidth}
\centering
\small
\setlength{\tabcolsep}{1.8pt}
\vspace{-2em}
\caption{
Stage-wise ablation of RePrior. 
}
\label{tab:stage_ablation}

\begin{tabular}{@{}lcccc@{}}
\toprule
Configuration
& \shortstack{3D\\PSNR$\uparrow$}
& \shortstack{3D\\SSIM$\uparrow$}
& \shortstack{2D\\PSNR$\uparrow$}
& \shortstack{2D\\SSIM$\uparrow$}
\\
\midrule

Prior-R$^2$-GS
& 32.03
& 0.8966
& 46.46
& 0.9956
\\

+ SCP
& \textbf{32.58}
& \textbf{0.9033}
& 46.30
& 0.9952
\\

+ MCR
& 32.24
& 0.8989
& 46.63
& \textbf{0.9957}
\\

+ HRE (Full)
& 32.43
& 0.9012
& \textbf{47.10}
& 0.9956
\\

\bottomrule
\end{tabular}
\end{wraptable}

HRE further expands the correction space around the measurement-feasible MCR
anchor. It increases 3D PSNR from 32.24 to 32.43~dB ($+0.19$~dB) and 3D SSIM
from 0.8989 to 0.9012, while observed-view projection PSNR increases from
46.63 to 47.10~dB ($+0.47$~dB). Projection SSIM remains nearly unchanged at
0.9956. These results show that the complementary voxel- and Gaussian-domain
residual freedoms recover additional useful corrections without sacrificing
the measurement feasibility established by MCR.
\section{Conclusion}
\label{sec:conclusion}

We propose RePrior, a multi-stage select--reconcile--enhance framework for
transferring learned priors into patient-specific Gaussian reconstruction for
sparse-view CBCT. RePrior identifies useful local prior corrections, reconciles
them under patient-specific X-ray measurements, and recovers residual
information beyond the restricted reconciliation space. This design separates
correction selection from measurement enforcement while preserving additional
freedom for reconstruction refinement. Across LUNA16, PANORAMA, and PENGWIN,
RePrior improves 3D PSNR by approximately 2.6~dB over R$^2$-Gaussian and
provides further volumetric and projection-fidelity gains beyond matched
prior-assisted reconstruction, with consistent improvements in held-out-view
fidelity. These results support using learned priors as sources of selective,
measurement-compatible corrections rather than transferring them
unconditionally as reconstruction targets.

\begingroup
\small
\setlength{\bibsep}{1pt}
\bibliographystyle{reprior}
\bibliography{references}
\endgroup
\clearpage
\appendix
\section*{APPENDIX}
\addcontentsline{toc}{section}{APPENDIX}
\section{Detailed Prior-Assisted \texorpdfstring{R$^2$-GS}{R2-GS} Reconstruction}
\label{app:prior_r2gs}

This section provides the implementation details of the prior-assisted
R$^2$-GS reconstruction used to construct the reference volume $V_R$ in
Sec.~\ref{sec:method_overview}.
We first obtain a frozen learned volumetric prior, establish a
measurement-driven Gaussian reconstruction prefix, and only then introduce
full-volume spatial prior supervision.
All per-case reconstruction experiments are conducted on NVIDIA V100 GPUs.

\subsection{Learned Volumetric Prior}
\label{app:prior_generation}

Given the 25 observed sparse-view X-ray projections $Y_{25}$, we use an
anatomy-specific DeepSparse-25 reconstructor $F_P$ to obtain the learned
volumetric prior
\begin{equation}
    V_P = F_P(Y_{25}).
    \label{eq:app_prior}
\end{equation}

\paragraph{DeepSparse-25 prior models.}
For the matched 25-view setting, we finetune anatomy-specific DeepSparse
models from the official pretrained weights
(\texttt{pretrain/ep\_700.pth}) following the official two-stage finetuning
protocol.
For the three anatomies evaluated in this paper, the finetuning sets contain
250 LUNA16 cases, 250 PANORAMA cases, and all 60 available PENGWIN training
cases, respectively.
Both stages are trained for 400 epochs.
Stage~2 is initialized from the corresponding Stage~1 epoch-400 checkpoint
and uses random sparse-view sampling with \texttt{min\_views=25}.
The final Stage~2 epoch-400 checkpoint is used throughout our experiments.

The same frozen DeepSparse-25 checkpoint serves two roles.
Its direct volumetric prediction is reported as the DeepSparse-25 baseline in
Table~\ref{tab:main_results}, while the identical prediction is used as the
learned prior $V_P$ for RePrior.
Thus, the DeepSparse-25 baseline and the prior used by RePrior share the same
checkpoint and acquisition setting.
No DeepSparse parameters are updated during patient-specific RePrior
reconstruction.

For RePrior inference, the 25 observed projections follow the fixed
25-view acquisition protocol used throughout our experiments.
The frozen network produces a $256^3$ normalized attenuation volume, which is
clipped to $[0,1]$.
Thus,
\begin{equation}
    V_P \in [0,1]^{256\times256\times256}.
\end{equation}
The prior network remains frozen for all test patients and is never adapted
using patient-specific ground truth. All subjects used for DeepSparse-25 fine-tuning are disjoint from the subjects
used in the reported evaluation.

\subsection{Measurement-Driven Gaussian Prefix}
\label{app:r2gs_prefix}

We first optimize R$^2$-GS using only the measured X-ray projections.
Let $\mathcal{G}$ denote the current Gaussian representation and
$\mathcal{V}(\mathcal{G})$ its voxelized attenuation volume.
Before prior supervision is activated, the objective is
\begin{equation}
\begin{aligned}
    \mathcal{L}_{\mathrm{base}}
    =
    \mathcal{L}_{\mathrm{L1,proj}}
    + 0.25\left(1-\mathrm{SSIM}\right)
    + 0.05\,\mathcal{L}_{\mathrm{TV}},
\end{aligned}
\label{eq:app_r2gs_base}
\end{equation}
where $\mathcal{L}_{\mathrm{TV}}$ is evaluated on a randomly sampled
$32^3$ voxelized subvolume.

The measurement-driven prefix is optimized until iteration $10{,}000$:
\begin{equation}
    t \leq 10{,}000
    \quad\Longrightarrow\quad
    \mathcal{L}^{(t)} = \mathcal{L}_{\mathrm{base}}.
\end{equation}
No learned-prior loss is applied during this prefix.
The current production configuration uses an FDK-based initialization with
approximately $50$K Gaussian primitives.
Densification follows the standard R$^2$-GS schedule during the initial
measurement-driven optimization and is disabled for the subsequent
prior-assisted continuation.

We use this staged schedule as the fixed reconstruction protocol throughout
our experiments.
We do not claim that the $10$K transition point is theoretically optimal,
as a matched prior-from-initialization ablation is not part of the frozen
experimental protocol.

\subsection{Full-Volume Spatial Prior Supervision}
\label{app:r2gs_prior_loss}

Starting from the measurement-driven prefix, we introduce full-volume spatial
supervision from $V_P$.
Define
\begin{equation}
    e = \mathcal{V}(\mathcal{G}) - V_P.
    \label{eq:app_prior_error}
\end{equation}
For a three-dimensional volume $x$, we define the mean first-order spatial
difference as
\begin{equation}
    \operatorname{FD}(x)
    =
    \frac{
        \displaystyle
        \sum_{a=1}^{3}
        \sum_{i:\,i+\mathbf{e}_a\in\Omega}
        \left|
            x_{i+\mathbf{e}_a}-x_i
        \right|
    }{
        \displaystyle
        \sum_{a=1}^{3} N_a
    },
    \label{eq:app_fd}
\end{equation}
where $N_a$ is the number of valid neighboring voxel pairs along axis $a$.

The spatial prior loss is
\begin{equation}
\begin{aligned}
    \mathcal{L}_{\mathrm{prior}}
    ={}&
    \frac{
        \left\langle
        \sqrt{e^2+10^{-8}}-10^{-4}
        \right\rangle
    }{
        \max\left(
            \langle |V_P|\rangle,
            10^{-3}
        \right)
    }
    \\
    &+
    0.25\,
    \frac{
        \operatorname{FD}(e)
    }{
        \max\left(
            \operatorname{FD}(V_P),
            10^{-3}
        \right)
    }.
\end{aligned}
\label{eq:app_prior_loss}
\end{equation}
The first term is a normalized Charbonnier volumetric discrepancy, whereas
the second term constrains first-order spatial variations.

The complete prior-assisted objective is
\begin{equation}
\begin{aligned}
    \mathcal{L}_{\mathrm{R2GS}}^{(t)}
    ={}&
    \mathcal{L}_{\mathrm{L1,proj}}
    +0.25(1-\mathrm{SSIM})
    +0.05\mathcal{L}_{\mathrm{TV}}
    \\
    &+
    \lambda_P\,
    \mathbf{1}_{P}(t)
    \mathcal{L}_{\mathrm{prior}},
\end{aligned}
\label{eq:app_r2gs_total}
\end{equation}
with
\begin{equation}
    \lambda_P = 0.19356790574571794.
\end{equation}
The prior term is evaluated on the complete $256^3$ volume.

\subsection{Prior Schedule and Reference Reconstruction}
\label{app:r2gs_schedule}

Full-volume voxelization is not performed at every optimization iteration.
The prior indicator in Eq.~(\ref{eq:app_r2gs_total}) is
\begin{equation}
    \mathbf{1}_{P}(t)
    =
    \mathbf{1}
    \left[
        t>10{,}000
        \ \wedge\
        t\bmod q_t=0
    \right],
\end{equation}
where
\begin{equation}
    q_t=
    \begin{cases}
        10, & 10{,}000<t<15{,}000,\\
        50, & t\geq15{,}000.
    \end{cases}
    \label{eq:app_prior_schedule}
\end{equation}
The first active prior evaluation therefore occurs at iteration $10{,}010$.
The prior weight is not decayed during continuation, and $V_P$ remains fixed.

The matched prior-assisted reference is taken at iteration $20{,}000$:
\begin{equation}
    \mathcal{G}_R
    =
    \mathcal{G}^{(20K)},
    \qquad
    V_R
    =
    \mathcal{V}(\mathcal{G}_R).
    \label{eq:app_reference}
\end{equation}
This P20 state is the reference reconstruction used by the subsequent SCP,
MCR, and HRE stages.
The same fixed Gaussian state $\mathcal G_R$ also serves as the
pre-correction measurement reference in the subsequent reconciliation
and residual-enhancement stages.

\section{Details of Selective Correction Proposal}
\label{app:scp}

SCP converts the remaining discrepancy between the learned prior and the
prior-assisted Gaussian reconstruction into an actionable signed correction.
This section specifies the training target, discrepancy representation,
FusionCNN architecture, and inference-time aggregation used in
Sec.~\ref{sec:scp}.

\subsection{Local Correction Actionability}
\label{app:scp_actionability}

Let
\begin{equation}
    d = V_P-V_R
\end{equation}
denote the remaining prior discrepancy, and let
\begin{equation}
    r_{\mathrm{GT}}
    =
    V_{\mathrm{GT}}-V_R
\end{equation}
denote the ground-truth-indicated reconstruction residual available only
during training.

For a local patch $P_k$ and body support $S$, we define their local alignment
as
\begin{equation}
    s_k
    =
    \sum_{i\in P_k\cap S}
    d_i\,r_{\mathrm{GT},i}.
    \label{eq:app_scp_alignment}
\end{equation}
A local prior direction is labeled actionable whenever it has positive
alignment with the ground-truth residual:
\begin{equation}
    a_k^*
    =
    \mathbf{1}[s_k>0].
    \label{eq:app_scp_gate_target}
\end{equation}

We additionally define the least-squares coefficient of the ground-truth
residual along the local prior-discrepancy direction:
\begin{equation}
    \beta_k^*
    =
    \frac{
        \displaystyle
        \sum_{i\in P_k\cap S}
        d_i\,r_{\mathrm{GT},i}
    }{
        \displaystyle
        \sum_{i\in P_k\cap S}
        d_i^2
        +10^{-30}
    }.
    \label{eq:app_scp_beta_target}
\end{equation}
Hence SCP learns two complementary quantities: whether a local prior
direction should be acted upon, and how strongly that direction should be
retained.

Importantly, the regression target $\beta_k^*$ is neither clipped nor masked
by $a_k^*$ during training.
The sign of the final correction is inherited from $V_P-V_R$, whereas the
predicted scalar controls its magnitude.

\subsection{Local Discrepancy Representation}
\label{app:scp_features}

The $256^3$ lattice is tiled with overlapping $16^3$ target patches using a
stride of $8$ voxels.
For each target patch, SCP extracts a $24^3$ spatial context using a
four-voxel halo.
A patch is retained if at least $25\%$ of its voxels belong to the
source-derived body support.

The spatial input contains three raw channels,
\begin{equation}
    X_k^{\mathrm{spa}}
    =
    \left[
        V_P,\,
        V_R,\,
        V_P-V_R
    \right]_{24^3},
    \label{eq:app_scp_spatial_input}
\end{equation}
together with a $75$-dimensional scalar descriptor
\begin{equation}
    \Phi_k
    \in
    \mathbb{R}^{75}.
\end{equation}

The descriptor summarizes multi-scale frequency structure, differential
morphology, local structural statistics, direct prior--reconstruction
disagreement, discrepancy-frequency structure, and frozen structural-anchor
relations.
For either source volume $v\in\{V_P,V_R\}$, we construct the Gaussian pyramid
\begin{equation}
    L_0=v,
    \qquad
    L_j=G_{\sigma_j}*v,
    \qquad
    \sigma_j\in\{1,2,4,8\},
\end{equation}
and define five frequency bands
\begin{equation}
\begin{aligned}
    B_1 &= L_0-L_1,\\
    B_2 &= L_1-L_2,\\
    B_3 &= L_2-L_3,\\
    B_4 &= L_3-L_4,\\
    B_5 &= L_4.
\end{aligned}
\label{eq:app_scp_bands}
\end{equation}
Their local energies are
\begin{equation}
    E_j=G_1*(B_j^2),
    \qquad
    T=\sum_{j=1}^{5}E_j.
    \label{eq:app_scp_energy}
\end{equation}
These quantities are used to construct band-energy ratios, spectral
centroids, high-to-low and mid-to-total ratios, and total spectral energy.

The complete feature groups are summarized below.

\begin{table}[t]
\centering
\small
\caption{
Composition of the $75$-dimensional SCP discrepancy descriptor.
All features are available at test time; ground truth is used only to build
the training targets in Eqs.~(\ref{eq:app_scp_gate_target}) and
(\ref{eq:app_scp_beta_target}).
}
\label{tab:app_scp_features}
\begin{tabular}{p{0.29\linewidth} c p{0.55\linewidth}}
\hline
Feature family & Dim. & Main quantities \\
\hline
Multi-scale frequency
& 18
& Five band-energy ratios, spectral centroid, high/low ratio,
mid/total ratio, and total energy for $V_P$ and $V_R$ \\

Differential morphology
& 38
& Gradient magnitude, surface normal, Hessian eigenstructure,
structure-tensor descriptors, local variance, contrast,
Laplacian, and boundary distance \\

Local attenuation
& 2
& Local attenuation values of $V_P$ and $V_R$ \\

Direct disagreement
& 5
& Absolute and relative attenuation discrepancy, gradient
difference, normal disagreement, and local NCC \\

Discrepancy spectrum
& 7
& Five discrepancy-band ratios, spectral centroid, and total
discrepancy energy \\

Structural-anchor relation
& 5
& Structural-feature distance and frozen
AGREE/DISAGREE/P1-ONLY/P2-ONLY indicators \\
\hline
Total & 75 & \\
\hline
\end{tabular}
\end{table}

For example, the relative attenuation discrepancy is
\begin{equation}
    d_{\mathrm{rel}}
    =
    \frac{
        |V_P-V_R|
    }{
        \max\left(
            0.5(|V_P|+|V_R|),
            10^{-4}
        \right)
    }.
    \label{eq:app_scp_rel_disagreement}
\end{equation}
Most non-negative magnitude descriptors are normalized using a casewise
$95$th-percentile scale within the body support and clipped to $[0,1]$.
Signed normals, bounded structural ratios, and NCC retain their natural
ranges.
Voxelwise feature fields are averaged over $P_k\cap S$, after which the
resulting $75$-dimensional vector is imputed and standardized using frozen
training-set statistics.

The structural-anchor descriptors are computed from frozen source-side
correspondence information.
No patient-specific observed-ray response is used by SCP; such measurement
reasoning is deferred to MCR.

\subsection{FusionCNN and Training Objective}
\label{app:scp_network}

SCP uses a lightweight dual-stream FusionCNN.
The spatial branch processes
$X_k^{\mathrm{spa}}\in\mathbb{R}^{3\times24\times24\times24}$
using three $3\times3\times3$ convolutional stages,
\begin{equation}
    3\rightarrow16\rightarrow32\rightarrow64,
\end{equation}
with GroupNorm and ReLU activations.
Two $2\times2\times2$ max-pooling operations reduce the spatial resolution,
and global average pooling produces a $64$-dimensional representation.

The scalar-feature branch maps the $75$-dimensional descriptor through
\begin{equation}
    75\rightarrow64\rightarrow32.
\end{equation}
The resulting $64$- and $32$-dimensional representations are concatenated
and passed through a $96\rightarrow64$ fusion layer.
Two scalar heads then predict a gate logit and a correction coefficient:
\begin{equation}
    (\hat a_k,\hat\beta_k)
    =
    F_{\mathrm{SCP}}
    \left(
        X_k^{\mathrm{spa}},
        \Phi_k
    \right).
    \label{eq:app_scp_network}
\end{equation}
The complete network contains approximately $84{,}034$ trainable parameters.

The training objective is
\begin{equation}
\begin{aligned}
    \mathcal{L}_{\mathrm{SCP}}
    ={}&
    \operatorname{BCEWithLogits}
    \left(
        \hat a_k,
        a_k^*;
        w_+
    \right)
    \\
    &+
    \operatorname{MSE}
    \left(
        \hat\beta_k,
        \beta_k^*
    \right),
\end{aligned}
\label{eq:app_scp_loss}
\end{equation}
where $w_+$ compensates for the positive/negative patch imbalance.

We train the SCP predictor with AdamW, learning rate $10^{-3}$,
weight decay $10^{-4}$, batch size $64$, and at most $30$ epochs with early
stopping.
The frozen checkpoint is trained on a multi-anatomy cohort of $155$ cases,
with an additional $20$ cases used for validation.
Training and validation subjects are separated from the reported evaluation
subjects.
The same frozen SCP checkpoint is used for every test patient without
case-specific fine-tuning.
Ground truth is used only to construct the offline supervision targets
$a_k^*$ and $\beta_k^*$ during SCP training and is never accessed during
test-time reconstruction.

\subsection{Inference and Overlap Aggregation}
\label{app:scp_inference}

At inference time, the predicted patch coefficient is
\begin{equation}
    b_k
    =
    \mathbf{1}
    \left[
        \sigma(\hat a_k)\geq0.5
    \right]
    \operatorname{clip}
    \left(
        \hat\beta_k,0,1
    \right).
    \label{eq:app_scp_inference_coeff}
\end{equation}
For each voxel $i$, coefficients from all overlapping patches are uniformly
averaged:
\begin{equation}
    \bar b_i
    =
    S_i
    \frac{
        \displaystyle
        \sum_{k:\,i\in P_k} b_k
    }{
        \displaystyle
        |\{k:i\in P_k\}|
    }.
    \label{eq:app_scp_overlap}
\end{equation}
No additional taper or center weighting is used.

The final SCP correction is
\begin{equation}
    D_A
    =
    \bar b
    \odot
    (V_P-V_R),
    \qquad
    V_A=V_R+D_A.
    \label{eq:app_scp_output}
\end{equation}
The support of $D_A$ specifies where a correction is proposed, its sign is
inherited from the learned-prior discrepancy, and its magnitude is controlled
by the predicted actionable coefficient.

\section{Details of Measurement-Constrained Reconciliation}
\label{app:mcr}

The main paper describes MCR through the abstract optimization
\begin{equation}
    \min_{\mathbf c}
    \|D_B(\mathbf c)-D_A\|_2^2
    \quad
    \mathrm{s.t.}\quad
    \mathcal{C}_{\mathrm{meas}}
    (r_0-H\mathbf c)\leq0,
    \qquad
    \mathbf c\in\mathcal C_{\mathrm{phys}}.
    \label{eq:app_mcr_abstract}
\end{equation}
This section specifies how the structured correction
$D_B(\mathbf c)$, the patient-specific response matrix $H$, and the
measurement constraints are instantiated in the production system.

MCR first establishes a coarse patient-specific, measurement-aware correction,
then parameterizes the remaining SCP correction using regional recovery and
signed compensation modes.
The latter are the variable modes represented by $\mathbf c$ in
Eq.~(\ref{eq:app_mcr_abstract}).

\subsection{Patient-Specific Measurement Representation}
\label{app:mcr_measurement}

Let $N$ denote the native 25-view prediction of the fixed
Prior-R$^2$-GS reference state $\mathcal G_R$, and let $Y$ denote the
corresponding observed measurements.
The native Gaussian prediction is used as the measurement baseline.

Incremental volumetric corrections are evaluated using a calibrated linear
TIGRE projector $P_{\mathrm{cal}}$.
The initial native residual is
\begin{equation}
    r = Y-N.
    \label{eq:app_mcr_native_residual}
\end{equation}
We deliberately distinguish the native Gaussian prediction $N$ from a
voxelized-volume replay: in general,
\begin{equation}
    N \neq P_{\mathrm{cal}}(V_R).
\end{equation}
Instead, $P_{\mathrm{cal}}$ is used to model the incremental response induced
by the volumetric correction modes.

\subsection{Coarse Measurement-Aware Correction}
\label{app:mcr_d0}

We first partition the $256^3$ lattice into $64$ non-overlapping
$64^3$ blocks $Q_j$.
Within block $Q_j$, define the local SCP correction
\begin{equation}
    d_j
    =
    \mathbf{1}_{Q_j}D_A.
\end{equation}
When $\|d_j\|_2>0$, the normalized desired direction is
\begin{equation}
    q_j
    =
    \frac{d_j}{\|d_j\|_2}.
    \label{eq:app_mcr_q}
\end{equation}

Let
\begin{equation}
    m_j=\mathbf{1}_{Q_j}S
\end{equation}
denote the corresponding body-support direction.
To introduce a coarse compensation degree of freedom that is distinct from
$q_j$, we orthogonalize $m_j$ with respect to $q_j$:
\begin{equation}
    \tilde u_j
    =
    m_j-\langle q_j,m_j\rangle q_j,
\end{equation}
and, whenever $\|\tilde u_j\|_2>0$,
\begin{equation}
    u_j
    =
    \frac{\tilde u_j}{\|\tilde u_j\|_2}.
    \label{eq:app_mcr_u}
\end{equation}

The target coefficient for a desired direction is
$\|d_j\|_2$, whereas the target coefficient for its orthogonal body direction
is zero.
Let $x^\star$ denote this target coefficient vector, and let $H_0$ collect
the calibrated X-ray responses of the corresponding coarse modes.
For a fixed regularization value $\lambda$, the coarse coefficients are
obtained from
\begin{equation}
    x_0(\lambda)
    =
    \arg\min_x
    \frac12\|x-x^\star\|_2^2
    +
    \frac{\lambda}{2s^2}
    \|r-H_0x\|_2^2,
    \label{eq:app_mcr_d0_ridge}
\end{equation}
where
\begin{equation}
    s^2
    =
    \lambda_{\max}(H_0^\top H_0).
\end{equation}
Equivalently, $x_0(\lambda)$ satisfies
\begin{equation}
    \left(
        I+
        \frac{\lambda}{s^2}
        H_0^\top H_0
    \right)x
    =
    x^\star
    +
    \frac{\lambda}{s^2}
    H_0^\top r.
    \label{eq:app_mcr_d0_linear}
\end{equation}

A frozen ridge ladder followed by direct projection checks selects a
measurement-compatible coefficient configuration.
The resulting direction is additionally restricted along an exact feasible
scalar interval before it is used by the subsequent regional stage.
We denote the resulting coarse correction by
\begin{equation}
    D_0
    =
    \sum_j
    x_j^{\mathrm{des}}q_j
    +
    \sum_j
    x_j^{\mathrm{body}}u_j.
    \label{eq:app_mcr_d0}
\end{equation}
Thus $D_0$ is a patient-specific, measurement-aware coarse realization of
the SCP proposal rather than a learned reconstruction module.

\subsection{Regional Recovery Modes}
\label{app:mcr_recovery}

After establishing $D_0$, we define the remaining SCP correction as
\begin{equation}
    E_A
    =
    D_A-D_0.
    \label{eq:app_mcr_ea}
\end{equation}
The volume is partitioned into $512$ non-overlapping $32^3$ regions $R_k$.
A region is active when its residual correction has non-negligible norm.

For an active region,
\begin{equation}
    E_{A,k}
    =
    \mathbf{1}_{R_k}E_A,
    \qquad
    n_k
    =
    \|E_{A,k}\|_2.
\end{equation}
We construct the unit-$\ell_2$ recovery basis
\begin{equation}
    \psi_k
    =
    \frac{
        \mathbf{1}_{R_k}E_A
    }{
        n_k
    },
    \label{eq:app_mcr_recovery_mode}
\end{equation}
with coefficient constraint
\begin{equation}
    0\leq x_k\leq n_k.
    \label{eq:app_mcr_recovery_bound}
\end{equation}
The full recovery of region $R_k$ is therefore obtained at
\begin{equation}
    x_k^{\mathrm{full}}=n_k,
    \qquad
    n_k\psi_k
    =
    \mathbf{1}_{R_k}E_A.
    \label{eq:app_mcr_full_recovery}
\end{equation}

This normalization separates the response-space basis direction from the
actual magnitude of the proposed regional correction.

\subsection{Signed Compensation Modes}
\label{app:mcr_compensation}

Regional recovery modes only preserve or suppress the residual SCP direction.
To provide limited attenuation-redistribution freedom under the X-ray
constraints, we additionally construct signed compensation modes.

Within an active region, define the positive and negative residual masses
\begin{equation}
    p(\mathbf x)
    =
    \max(E_A(\mathbf x),0)\,S(\mathbf x),
\end{equation}
\begin{equation}
    n(\mathbf x)
    =
    \max(-E_A(\mathbf x),0)\,S(\mathbf x).
\end{equation}
When both contain sufficient mass, their centroids are
\begin{equation}
    \mu_+
    =
    \frac{
        \sum_{\mathbf x}
        \mathbf x\,p(\mathbf x)
    }{
        \sum_{\mathbf x}p(\mathbf x)
    },
    \qquad
    \mu_-
    =
    \frac{
        \sum_{\mathbf x}
        \mathbf x\,n(\mathbf x)
    }{
        \sum_{\mathbf x}n(\mathbf x)
    }.
    \label{eq:app_mcr_centroids}
\end{equation}
We retain the pair only when
\begin{equation}
    \|\mu_+-\mu_-\|_2\geq4
\end{equation}
voxels.

We then construct two support-masked isotropic spatial Gaussian lobes
\begin{equation}
    g_{\pm}(\mathbf x)
    =
    \exp
    \left(
        -\frac{
            \|\mathbf x-\mu_{\pm}\|_2^2
        }{
            2\sigma^2
        }
    \right)
    S(\mathbf x),
    \qquad
    \sigma=6.
\end{equation}
The approximately mass-balanced signed mode is
\begin{equation}
    \psi_j^{\mathrm{pair}}
    =
    \frac{g_+}{\sum_{\mathbf x}g_+(\mathbf x)}
    -
    \frac{g_-}{\sum_{\mathbf x}g_-(\mathbf x)}.
    \label{eq:app_mcr_pair}
\end{equation}

These Gaussian lobes are purely spatial basis functions and are unrelated to
the R$^2$-GS Gaussian primitives optimized in HRE.

To bound the compensation amplitude, let
\begin{equation}
    f
    =
    \min(V_R+D_A,\;V_R+D_0)
\end{equation}
elementwise, and define the global attenuation floor
\begin{equation}
    z_{\min}
    =
    \min_i(V_R+D_0)_i.
\end{equation}
The maximal admissible local amplitude implied by this floor is
\begin{equation}
    h
    =
    \min_{i:\psi^{\mathrm{pair}}_i\neq0}
    \frac{
        f_i-z_{\min}
    }{
        |\psi^{\mathrm{pair}}_i|
    }.
\end{equation}
Let
\begin{equation}
    m
    =
    \sum_{i\in R_k}
    \max(V_R(i),0)S_i.
\end{equation}
The pair amplitude is
\begin{equation}
    a
    =
    \min
    \left(
        0.20m,\;
        0.5\max(h,0)
    \right).
    \label{eq:app_mcr_pair_amp}
\end{equation}
We define
\begin{equation}
    \phi_j
    =
    a\psi_j^{\mathrm{pair}},
    \qquad
    b_j
    =
    \frac{\phi_j}{\|\phi_j\|_2},
\end{equation}
with signed coefficient bound
\begin{equation}
    -\|\phi_j\|_2
    \leq z_j
    \leq
    \|\phi_j\|_2.
    \label{eq:app_mcr_pair_bound}
\end{equation}

At most $256$ candidate pairs are constructed.
A frozen source-only response-space screening procedure retains at most $64$
compensation modes.
The purpose of these modes is to introduce controlled signed redistribution
freedom; we do not assume that an arbitrary pair of individually conflicting
corrections becomes feasible when combined.

\subsection{Joint Structured Reconciliation}
\label{app:mcr_joint}

Combining the coarse correction, regional recovery modes, and signed
compensation modes yields the raw structured correction family
\begin{equation}
\begin{aligned}
    D^{\mathrm{raw}}(x,z)
    =
    D_0
    +
    \sum_{k\in\mathcal K}
    x_k\psi_k
    +
    \sum_{j\in\mathcal J}
    z_jb_j.
\end{aligned}
\label{eq:app_mcr_structured}
\end{equation}
Let $B_m$ denote any variable recovery or compensation basis.
Its patient-specific X-ray response is
\begin{equation}
    h_m
    =
    P_{\mathrm{cal}}(B_m),
\end{equation}
and
\begin{equation}
    H
    =
    [h_1,\ldots,h_M].
    \label{eq:app_mcr_H}
\end{equation}

After the coarse correction, the measurement residual becomes
\begin{equation}
    r_0
    =
    Y-N-P_{\mathrm{cal}}(D_0).
    \label{eq:app_mcr_r0}
\end{equation}
For coefficient vector
\begin{equation}
    \mathbf c=[x;z],
\end{equation}
the predicted residual is
\begin{equation}
    r(\mathbf c)
    =
    r_0-H\mathbf c.
    \label{eq:app_mcr_residual}
\end{equation}

The exact production objective is
\begin{equation}
    \min_{\mathbf c}
    \frac{
        \left\|
            D_0+B\mathbf c-D_A
        \right\|_2^2
    }{
        2\|E_A\|_2^2
    },
    \label{eq:app_mcr_exact_objective}
\end{equation}
subject to group-wise measurement constraints
\begin{equation}
    \sum_{v\in G}
    \left\|
        r_{0,v}
        -
        H_v\mathbf c
    \right\|_2^2
    \leq
    T_G,
    \label{eq:app_mcr_group_guard}
\end{equation}
and the recovery and compensation coefficient bounds defined above.

The denominator in Eq.~(\ref{eq:app_mcr_exact_objective}) is constant for a
given patient and therefore does not alter the minimizer.
This exact formulation corresponds to the simpler unnormalized objective used
in the main paper.

The observed views are partitioned into the complete $25$-view set, a
$20$-view fitting subset, and a fixed $5$-view calibration subset.
Their SSE limits relative to the native reference are
\begin{equation}
\begin{aligned}
    T_{\mathrm{full25}}
    &=
    \mathrm{SSE}_{\mathrm{full25}}(N),
    \\
    T_{\mathrm{fit20}}
    &=
    \mathrm{SSE}_{\mathrm{fit20}}(N)
    10^{0.05/10},
    \\
    T_{\mathrm{cal5}}
    &=
    \mathrm{SSE}_{\mathrm{cal5}}(N)
    10^{0.05/10},
\end{aligned}
\label{eq:app_mcr_group_budgets}
\end{equation}
while each individual observed view satisfies
\begin{equation}
    \mathrm{SSE}_{v}
    \leq
    \mathrm{SSE}_{v}(N)
    10^{0.15/10}.
    \label{eq:app_mcr_view_budget}
\end{equation}

The regional and recovery-plus-compensation problems are solved with SLSQP,
using a Phase-I feasibility step followed by target minimization, at most
$1500$ iterations per start, and $\mathrm{ftol}=10^{-12}$.
All candidate solutions are subsequently replayed with the production
projector before being accepted.

\subsection{Physical Repair and Final Certification}
\label{app:mcr_repair}

The structured solve does not constrain every voxel of the raw candidate to
be non-negative.
Since $D_B$ is a signed correction, the physical requirement is instead
\begin{equation}
    V_B=V_R+D_B\geq0.
\end{equation}

Let
\begin{equation}
    V_B^{\mathrm{raw}}
    =
    V_R+D_B^{\mathrm{raw}}.
\end{equation}
We first construct the zero-clipped target
\begin{equation}
    V_{\mathrm{clip}}
    =
    \max(V_B^{\mathrm{raw}},0).
    \label{eq:app_mcr_clip}
\end{equation}
Because clipping changes the measurement response, this volume is not
accepted directly.
Starting from an already certified repaired coarse anchor $V_{B0}$, we search
along
\begin{equation}
    V(t)
    =
    V_{B0}
    +
    t\left(
        V_{\mathrm{clip}}-V_{B0}
    \right),
    \qquad
    0\leq t\leq1,
    \label{eq:app_mcr_repair_line}
\end{equation}
for the feasible point that best preserves the target correction.

Equivalently, in correction space,
\begin{equation}
    D_B
    =
    D_{B0}
    +
    t
    \left(
        D_{\mathrm{clip}}-D_{B0}
    \right).
    \label{eq:app_mcr_final_db}
\end{equation}
The repaired candidate is reprojected and must pass all inherited measurement
constraints, body-support checks, numerical-validity checks, and
non-negativity.
The resulting volume
\begin{equation}
    V_B
    =
    V_R+D_B
    \geq0
    \label{eq:app_mcr_vb}
\end{equation}
is the measurement-feasible anchor used by HRE.

\paragraph{Algorithmic summary.}
MCR can be summarized as follows:
\begin{enumerate}
    \item Construct a coarse $64^3$ measurement-aware correction $D_0$.
    \item Compute the unresolved SCP correction $E_A=D_A-D_0$.
    \item Construct normalized $32^3$ regional recovery modes.
    \item Construct and screen signed spatial compensation modes.
    \item Compute patient-specific X-ray responses for all variable modes.
    \item Jointly optimize their coefficients under group-wise and per-view
    measurement constraints.
    \item Replay candidate corrections with the production projector.
    \item Enforce non-negative attenuation through a measurement-feasible
    line repair and certify the final $V_B$.
\end{enumerate}

\section{Details of Hybrid Residual Enhancement}
\label{app:hre}

MCR deliberately establishes measurement feasibility in a restricted
structured correction space.
HRE subsequently introduces bounded voxel- and Gaussian-domain residual
freedom around this feasible anchor.
This section specifies the residual representation, local-response
optimization, measurement guards, and endpoint selection referenced in
Sec.~\ref{sec:hre}.

\subsection{Residual Correction and Gaussian Reference}
\label{app:hre_reference}

The correction realized by MCR is
\begin{equation}
    D_B=V_B-V_R.
\end{equation}
The remaining SCP correction is therefore
\begin{equation}
    R_B
    =
    D_A-D_B
    =
    V_A-V_B.
    \label{eq:app_hre_rb}
\end{equation}
We make no assumption that $R_B$ is measurement-feasible on its own.

The Gaussian reference used throughout HRE is the matched P20
Prior-R$^2$-GS checkpoint,
\begin{equation}
    \mathcal{G}_{\mathrm{ref}}
    =
    \mathcal{G}_R^{(20K)}.
    \label{eq:app_hre_gref}
\end{equation}
It is not a Gaussian representation refitted to $V_B$.
Its topology remains fixed throughout HRE, with no densification or pruning.

\subsection{Hybrid Voxel--Gaussian Residual Representation}
\label{app:hre_representation}

The voxel branch directly moves along the unresolved SCP direction:
\begin{equation}
    \Delta V_{\mathrm{vox}}(\alpha)
    =
    \alpha R_B,
    \qquad
    0\leq\alpha\leq1.
    \label{eq:app_hre_voxel}
\end{equation}

In parallel, let $\mathcal G_\theta$ denote the bounded Gaussian state obtained
from $\mathcal G_{\mathrm{ref}}$.
Because Gaussian parameter changes are nonlinear after voxelization, we
measure their actual volumetric contribution using the production voxelizer:
\begin{equation}
    \Delta V_G(\theta)
    =
    \mathcal V_{\mathrm{prod}}(\mathcal G_\theta)
    -
    \mathcal V_{\mathrm{prod}}(\mathcal G_{\mathrm{ref}}).
    \label{eq:app_hre_gaussian_residual}
\end{equation}

The hybrid candidate is therefore
\begin{equation}
    V_C(\alpha,\theta)
    =
    V_B
    +
    \alpha R_B
    +
    \Delta V_G(\theta).
    \label{eq:app_hre_candidate}
\end{equation}

The source target is the SCP endpoint $V_A$.
We define the normalized source objective
\begin{equation}
    \mathcal L_A(V)
    =
    \frac{
        \|V-V_A\|_2^2
    }{
        \|V_A-V_B\|_2^2
    }.
    \label{eq:app_hre_source_loss}
\end{equation}
Thus $\mathcal L_A(V_B)=1$.
HRE seeks additional source recovery while preserving the measurement
feasibility inherited from MCR.

\subsection{Certified Scalar Recovery}
\label{app:hre_scalar}

Before introducing Gaussian residual freedom, we first consider the pure
voxel path
\begin{equation}
    V(\alpha)
    =
    V_B+\alpha R_B,
    \qquad
    0\leq\alpha\leq1.
\end{equation}
We find the largest actually certified scalar coefficient $\alpha_\star$ and
define
\begin{equation}
    V_\star
    =
    V_B+\alpha_\star R_B.
    \label{eq:app_hre_vstar}
\end{equation}
The scalar search uses the quadratic measurement response only to obtain an
upper bound; final feasibility is determined by real forward projection and
bisection.
Consequently, $V_\star$ is selected without access to ground truth or held-out
views and becomes the starting anchor of the coupled voxel--Gaussian
refinement.

\subsection{Bounded Gaussian Residual Space}
\label{app:hre_bounds}

HRE allows bounded updates to Gaussian position, attenuation, scale, and
rotation while preserving the P20 topology.
The trust region relative to $\mathcal G_{\mathrm{ref}}$ is summarized in
Table~\ref{tab:app_hre_bounds}.

\begin{table}[t]
\centering
\small
\caption{Gaussian trust bounds used by HRE.}
\label{tab:app_hre_bounds}
\begin{tabular}{p{0.38\linewidth} p{0.50\linewidth}}
\hline
Quantity & Constraint \\
\hline
Position
& $\|\Delta x_i\|_2\leq0.2$ normalized scene units \\

Attenuation / density
& $\left|\log(\rho_i/\rho_i^{\mathrm{ref}})\right|\leq0.1$ \\

Per-axis scale
& $\left|\log(s_{i,a}/s_{i,a}^{\mathrm{ref}})\right|\leq0.1$ \\

Physical scale
& $0.001\leq s_{i,a}\leq1$ \\

Rotation
& geodesic change $\leq10^\circ$ \\

Topology
& fixed; no densification or pruning \\
\hline
\end{tabular}
\end{table}

These bounds allow the Gaussian branch to redistribute residual attenuation
through primitive geometry and support without reopening an unrestricted
patient-specific Gaussian reconstruction.

\subsection{Local-Response Optimization}
\label{app:hre_local_response}

Rather than jointly optimizing all Gaussian parameters with a conventional
fixed-learning-rate optimizer, HRE constructs a low-dimensional local
response space at every outer iteration.

Up to four high-shortfall $32^3$ regions contribute local density directions,
together with one global density direction and up to three regional geometry
directions.
Each direction is derived from the negative gradient of the source objective
with respect to the raw Gaussian parameters, with parameter-group scaling
inherited from the frozen P20 optimizer state.
After normalization, at most eight Gaussian directions are retained.

Let these directions be
$\{d_k\}_{k=1}^{K}$, $K\leq8$.
A local Gaussian update is represented as
\begin{equation}
    \Delta p
    =
    \sum_{k=1}^{K}
    \gamma_k d_k,
    \label{eq:app_hre_local_update}
\end{equation}
with
\begin{equation}
    \|\gamma\|_1\leq1.
    \label{eq:app_hre_l1}
\end{equation}
The voxel coefficient is simultaneously updated within
\begin{equation}
    |\Delta\alpha|
    \leq0.125\,\eta,
    \qquad
    0\leq\alpha\leq1,
    \label{eq:app_hre_alpha_step}
\end{equation}
where $\eta$ is the current trust scale.

HRE builds a local quadratic approximation of the source objective and of the
measurement responses over the voxel direction and the selected Gaussian
directions.
A bounded local QCQP proposes the joint coefficient update.

Importantly, this local solution is only a proposal.
Every candidate is subsequently reconstructed with the production voxelizer
and evaluated using the actual X-ray forward model.
The response approximation therefore improves search efficiency without
replacing the physical certification step.

\subsection{Measurement Guards}
\label{app:hre_guards}

HRE evaluates every real candidate under two calibrated projection scales,
\begin{equation}
    s\in\{\mathrm{cert},\mathrm{common}\}.
\end{equation}
The \emph{cert} scale uses the frozen per-view calibration employed for
certification, whereas the \emph{common} scale uses a case-specific global
scale fixed during preflight from the reference Gaussian prediction.

For scale $s$ and view group $G$, define
\begin{equation}
    S_{s,G}(V)
    =
    \sum_{v\in G}
    \left\|
        P_s(V)_v-Y_v
    \right\|_2^2.
    \label{eq:app_hre_sse}
\end{equation}

Let $S^{\mathrm{ref}}_{s,G}$ denote the corresponding SSE of the fixed
pre-correction reference prediction produced by $\mathcal G_R$ under
scale $s$.

For
\begin{equation}
    G\in
    \{
        \mathrm{full25},
        \mathrm{fit20},
        \mathrm{cal5}
    \},
\end{equation}
HRE first inherits the MCR-stage cap
\begin{equation}
    T_{s,G}^{B}
    =
    \min
    \left[
        S^{\mathrm{ref}}_{s,G}\rho_G,\;
        S_{s,G}(V_B)
        10^{2\times10^{-7}/10}
    \right],
    \label{eq:app_hre_B_cap}
\end{equation}
where
\begin{equation}
    \rho_{\mathrm{full25}}=1,
    \qquad
    \rho_{\mathrm{fit20}}
    =
    \rho_{\mathrm{cal5}}
    =
    10^{0.05/10}.
\end{equation}

After the certified scalar anchor $V_\star$ has been established, the coupled
HRE cap is tightened to
\begin{equation}
    T_{s,G}^{C}
    =
    \min
    \left[
        T_{s,G}^{B},\;
        S_{s,G}(V_\star)
        10^{2\times10^{-7}/10}
    \right].
    \label{eq:app_hre_C_cap}
\end{equation}
For every individual observed view,
\begin{equation}
    S_{s,v}(V_C)
    \leq
    S^{\mathrm{ref}}_{s,v}
    10^{0.15/10}.
    \label{eq:app_hre_view_guard}
\end{equation}
All pooled and per-view constraints must hold under both calibration scales.

These constraints instantiate the measurement guards referred to in
Sec.~\ref{sec:hre}: HRE is allowed to expand representation freedom only
within the measurement envelope already established by MCR and the certified
scalar recovery.

\subsection{Spatial and Physical Validity Constraints}
\label{app:hre_spatial_guard}

Let $O=\neg S$ denote the region outside the source-derived support, and define
the fixed normalization
\begin{equation}
    Z_O
    =
    \|V_R[O]\|_2.
\end{equation}
We constrain both the total outside-support deviation and the additional
deviation introduced after MCR:
\begin{equation}
    \frac{
        \|(V_C-V_R)[O]\|_2
    }{
        Z_O
    }
    \leq
    L_{\mathrm{total}},
    \label{eq:app_hre_out_total}
\end{equation}
and
\begin{equation}
    \frac{
        \|(V_C-V_B)[O]\|_2
    }{
        Z_O
    }
    \leq
    0.005.
    \label{eq:app_hre_out_increment}
\end{equation}
When $V_B$ already satisfies the $0.005$ total bound,
$L_{\mathrm{total}}=0.005$; otherwise the measured deviation of $V_B$ is
retained as the compatibility anchor.

Every candidate must additionally be finite and satisfy
\begin{equation}
    \min V_C \geq -10^{-7}.
\end{equation}
No post-hoc volume clipping is applied in HRE.

\subsection{Candidate Acceptance and Endpoint Selection}
\label{app:hre_selection}

For every local QCQP proposal, HRE evaluates the backtracking fractions
\begin{equation}
    1,\;
    \frac12,\;
    \frac14,\;
    \frac18,\;
    \frac1{16},\;
    \frac1{32},\;
    \frac1{64}.
    \label{eq:app_hre_backtracking}
\end{equation}
Each candidate is re-voxelized and reprojected.
It is accepted only if all Gaussian trust bounds, pooled measurement guards,
per-view constraints, spatial-support constraints, and numerical checks are
satisfied.

A candidate is accepted directly when the source objective improves by more
than $2\times10^{-7}$.
If the source objectives agree within this tolerance, the candidate is
accepted only when its common-scale full-25-view SSE improves beyond the
repeatability margin.

If a round is rejected, the previous certified Gaussian state is restored and
the trust scale is reduced.
The production configuration uses at most $30$ outer rounds, at most $100$
iterations for each local QCQP, $\mathrm{ftol}=10^{-12}$, and a
$1200$-second wall-time budget.
Optimization also terminates after six consecutive failed rounds or when the
trust scale becomes smaller than $1/64$.

All accepted states, including the scalar-only state $V_\star$, are certified
before endpoint selection.
Eligibility requires
\begin{equation}
    \mathcal L_A(V)
    \leq
    \mathcal L_A(V_\star)
    +
    2\times10^{-7}.
    \label{eq:app_hre_eligibility}
\end{equation}
Among eligible candidates, the final state is selected lexicographically by
\begin{equation}
    \left(
        \mathcal L_A,\;
        S_{\mathrm{common,full25}},\;
        \text{outer index}
    \right).
    \label{eq:app_hre_selection}
\end{equation}
Thus source fidelity is prioritized first, followed by common-scale
measurement consistency and a deterministic iteration tie-break.
If no coupled Gaussian update improves upon the certified scalar anchor,
$V_\star$ is retained.

The final reconstruction is the hybrid attenuation volume
\begin{equation}
    V^*
    =
    V_C(\alpha^*,\theta^*).
    \label{eq:app_hre_final}
\end{equation}
Both ground truth and held-out views remain excluded from HRE optimization
and endpoint selection.

\paragraph{Algorithmic summary.}
HRE can be summarized as follows:
\begin{enumerate}
    \item Compute $R_B=V_A-V_B$.
    \item Determine the largest measurement-certified scalar voxel recovery
    and obtain $V_\star$.
    \item Generate a small set of gradient-derived Gaussian residual
    directions.
    \item Build a local voxel--Gaussian response model and solve the bounded
    QCQP.
    \item Apply the proposed update using backtracking.
    \item Re-voxelize and reproject every real candidate.
    \item Accept only candidates that satisfy all measurement, support,
    numerical, and Gaussian trust constraints.
    \item Repeat until the stopping criterion is reached.
    \item Select the final source-first certified endpoint and output $V^*$.
\end{enumerate}


\clearpage
\setcounter{figure}{0}
\renewcommand{\thefigure}{S\arabic{figure}}
\renewcommand{\theHfigure}{S\arabic{figure}}

\section{Additional qualitative results}
\begin{figure*}[htbp]
    \centering
    \includegraphics[width=0.96\textwidth]
    {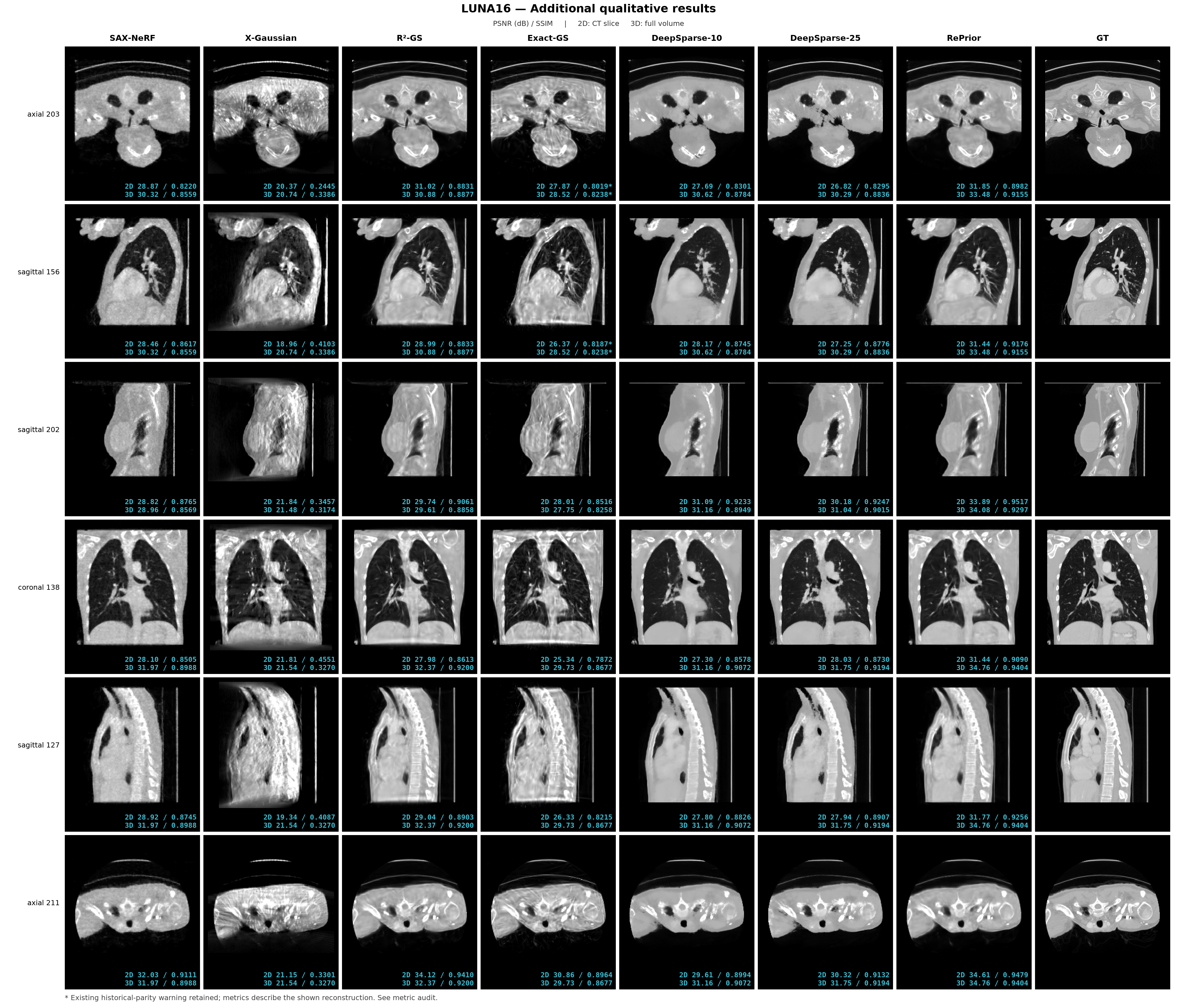}
    \caption{
    \textbf{Additional qualitative results on LUNA16 (Part I).}
    Each row shows an additional representative slice, while columns compare
    SAX-NeRF, X-Gaussian, R$^2$-GS, Exact-GS, DeepSparse-10,
    DeepSparse-25, RePrior, and GT.
    }
    \label{fig:app_luna_part1}
\end{figure*}

\begin{figure*}[p]
    \centering
    \includegraphics[width=0.96\textwidth]
    {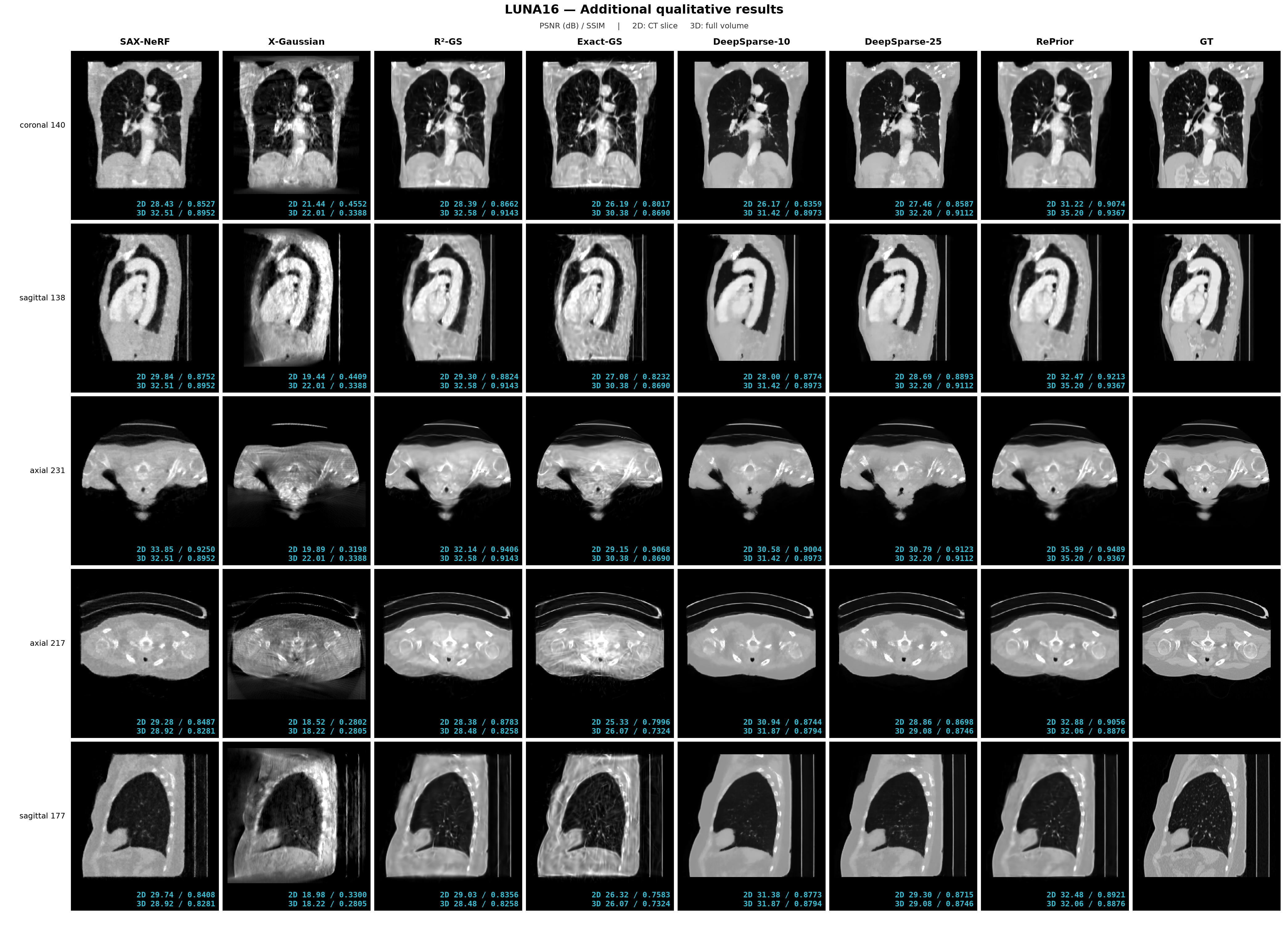}
    \caption{
    \textbf{Additional qualitative results on LUNA16 (Part II).}
    Additional slices and viewing planes using the same method ordering as in
    Fig.~\ref{fig:app_luna_part1}.
    }
    \label{fig:app_luna_part2}
\end{figure*}

\begin{figure*}[p]
    \centering
    \includegraphics[width=0.96\textwidth]
    {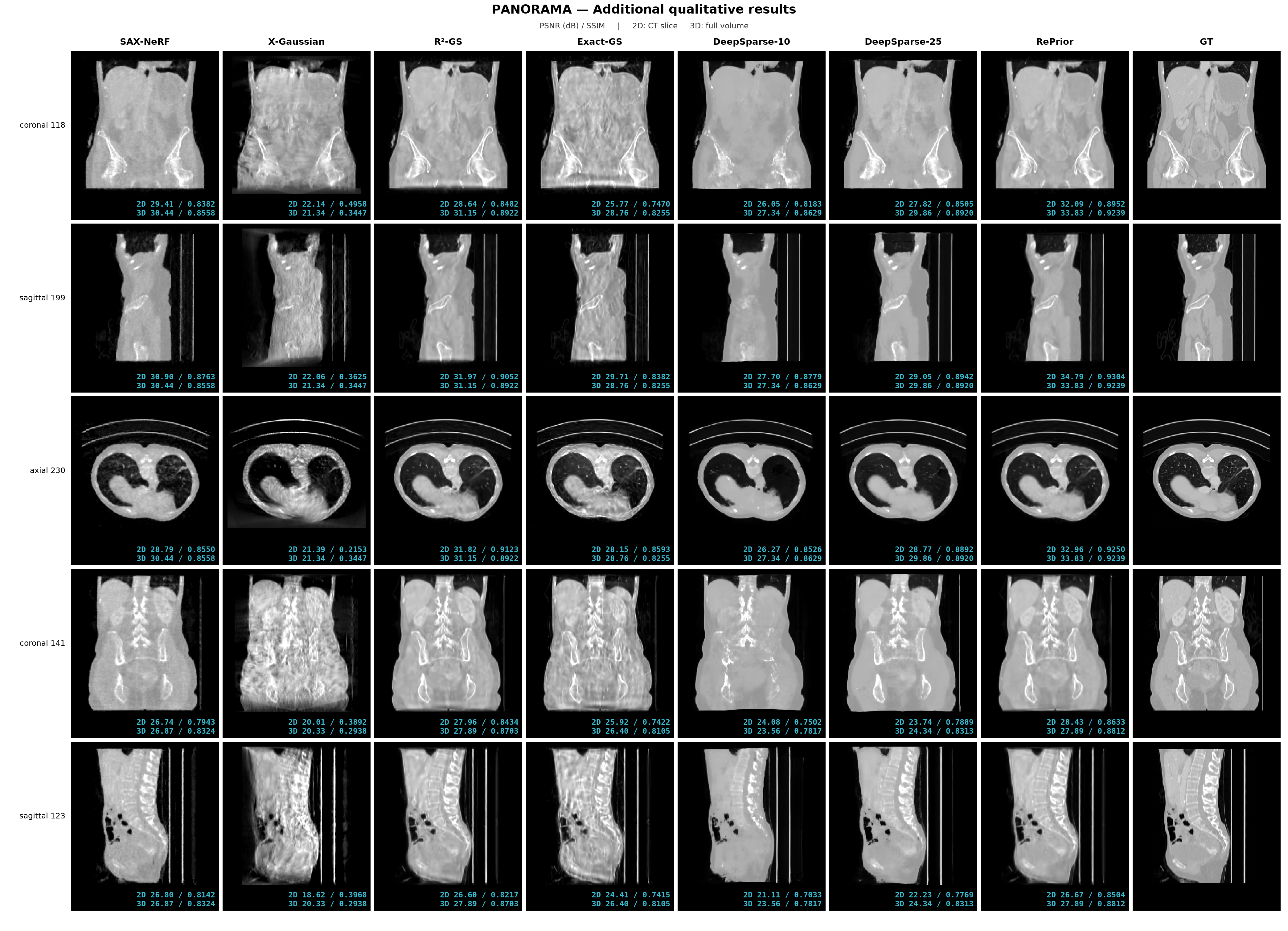}
    \caption{
    \textbf{Additional qualitative results on PANORAMA (Part I).}
    Each row shows an additional representative slice, while columns compare
    SAX-NeRF, X-Gaussian, R$^2$-GS, Exact-GS, DeepSparse-10,
    DeepSparse-25, RePrior, and GT.
    }
    \label{fig:app_panorama_part1}
\end{figure*}

\begin{figure*}[p]
    \centering
    \includegraphics[width=0.96\textwidth]
    {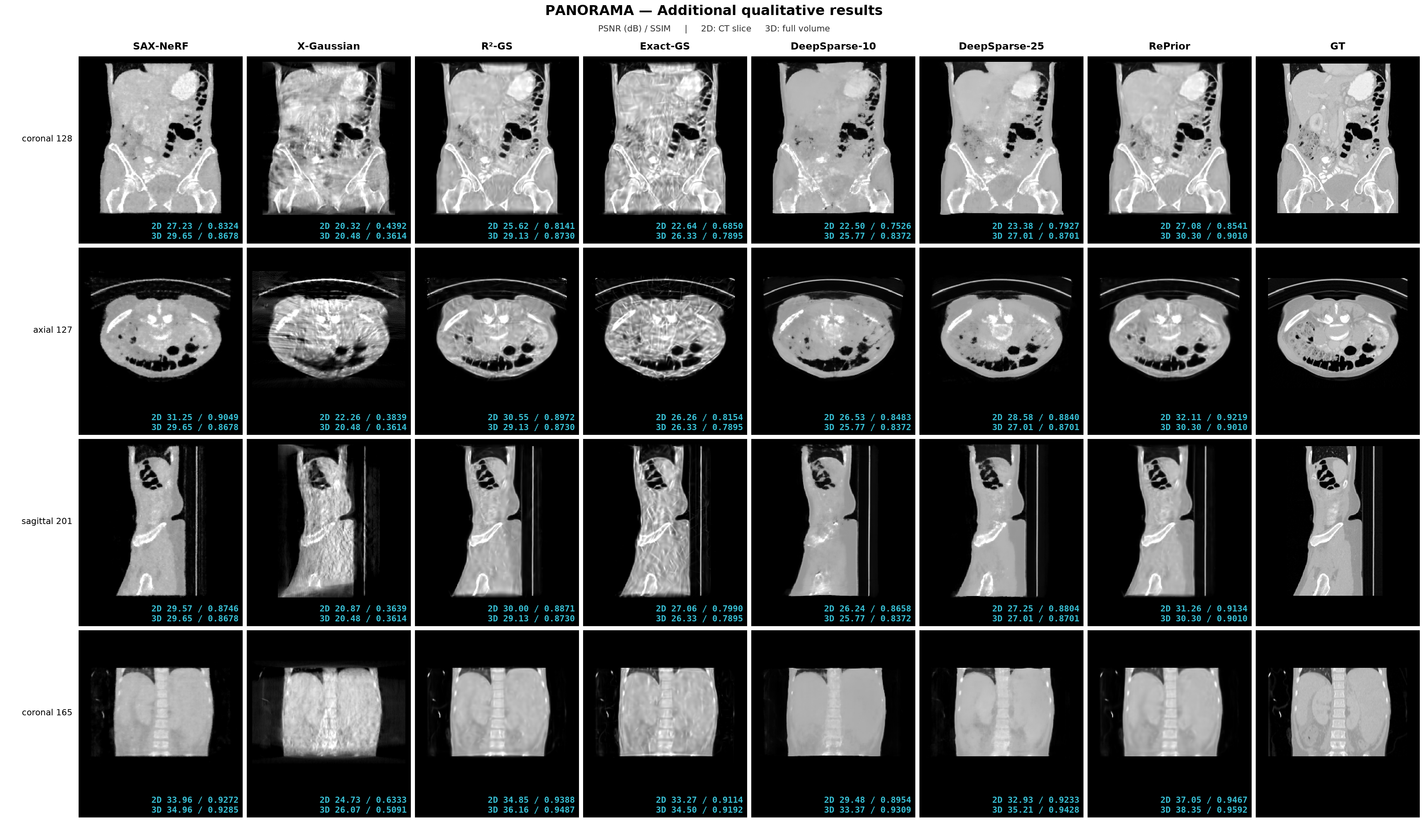}
    \caption{
    \textbf{Additional qualitative results on PANORAMA (Part II).}
    Additional slices and viewing planes using the same method ordering as in
    Fig.~\ref{fig:app_panorama_part1}.
    }
    \label{fig:app_panorama_part2}
\end{figure*}

\begin{figure*}[p]
    \centering
    \includegraphics[width=0.96\textwidth]
    {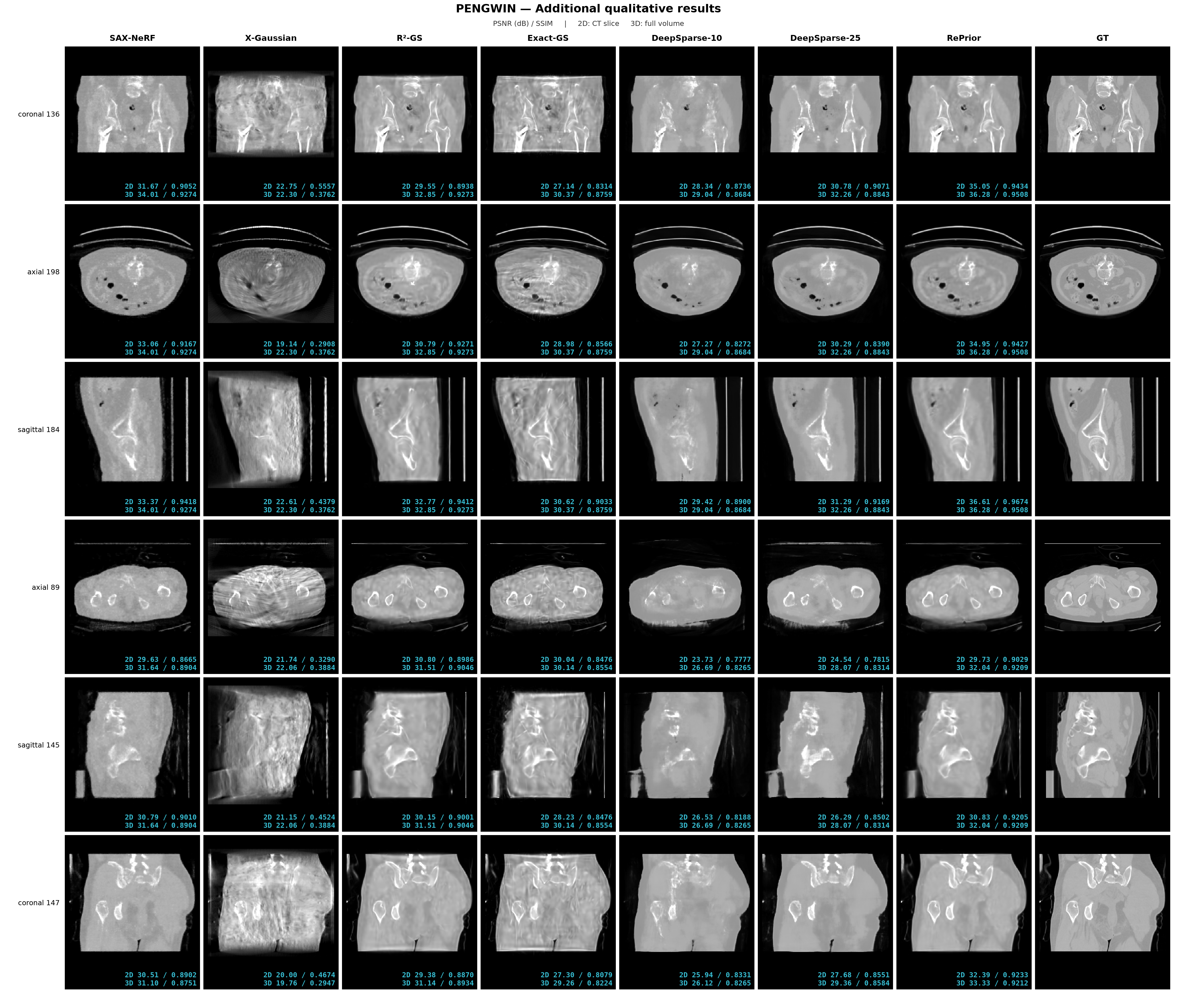}
    \caption{
    \textbf{Additional qualitative results on PENGWIN (Part I).}
    Each row shows an additional representative slice, while columns compare
    SAX-NeRF, X-Gaussian, R$^2$-GS, Exact-GS, DeepSparse-10,
    DeepSparse-25, RePrior, and GT.
    }
    \label{fig:app_pengwin_part1}
\end{figure*}

\begin{figure*}[p]
    \centering
    \includegraphics[width=0.96\textwidth]
    {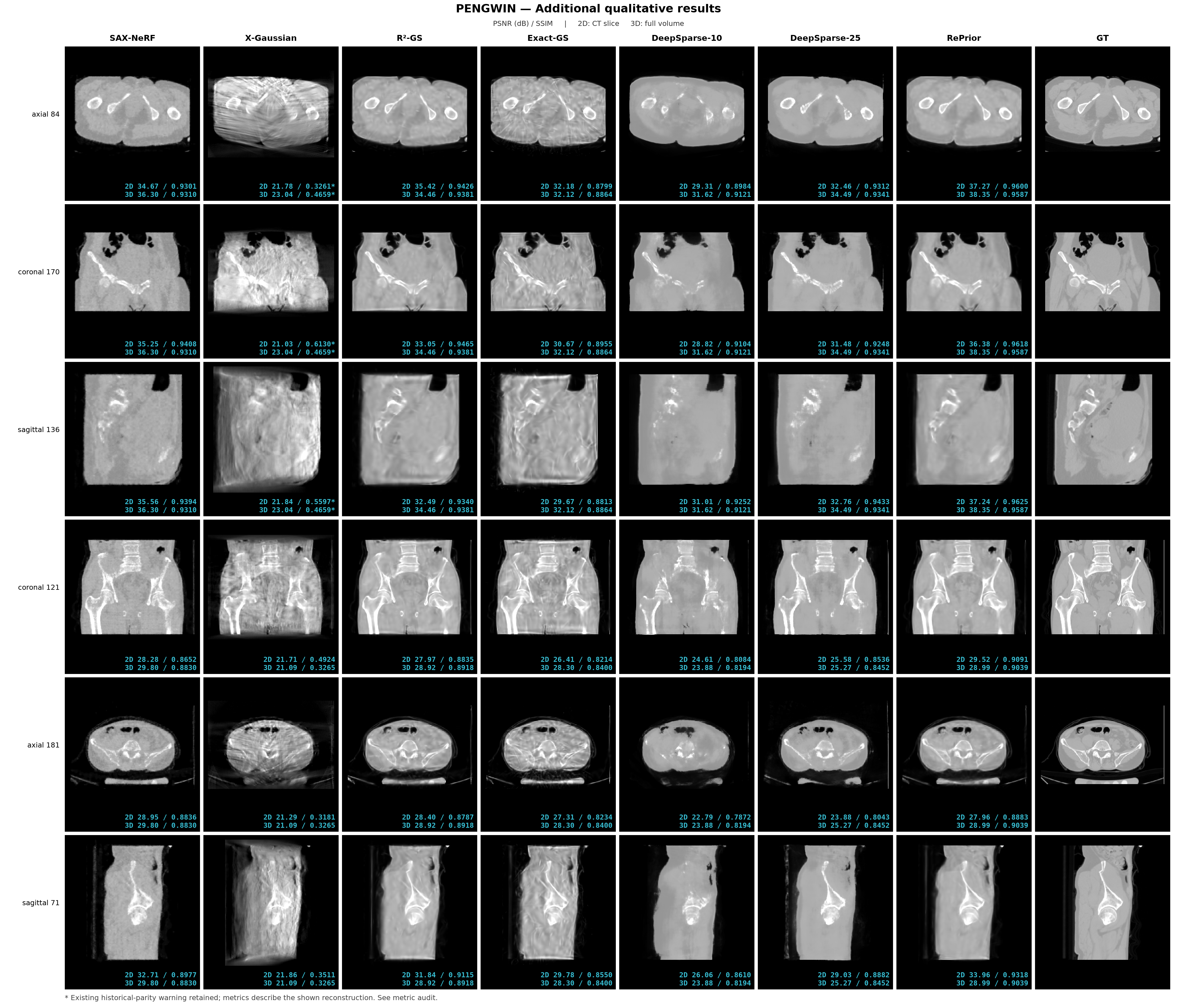}
    \caption{
    \textbf{Additional qualitative results on PENGWIN (Part II).}
    Additional slices and viewing planes using the same method ordering as in
    Fig.~\ref{fig:app_pengwin_part1}.
    }
    \label{fig:app_pengwin_part2}
\end{figure*}

To complement the quantitative results in the main paper, we provide additional qualitative comparisons on LUNA16, PANORAMA, and PENGWIN in Figs~\ref{fig:app_luna_part1}-~\ref{fig:app_pengwin_part2}.
The supplementary figures below show additional slices and viewing planes.
Columns correspond to SAX-NeRF, X-Gaussian, R$^2$-GS, Exact-GS,
DeepSparse-10, DeepSparse-25, RePrior, and the ground truth (GT).
\vspace{3em}

\clearpage

\end{document}